\documentclass[journal=jacsat,manuscript=letter]{achemso}

\usepackage{chemformula} 
\usepackage[T1]{fontenc} 
\usepackage{xcolor}
\usepackage[normalem]{ulem}
\usepackage{tikz}
\usetikzlibrary{shapes}
\usetikzlibrary{decorations.pathmorphing}
\usepackage{textcomp} 
\usepackage{array}
\usepackage{amsmath,amssymb}
\usepackage{gensymb}
\usepackage{siunitx}
\usepackage{tabularx}
\usepackage{multicol}
\usepackage{multirow}
\usepackage{caption}
\usepackage{achemso}
\usepackage{enumitem, pifont}  
\usepackage{xargs}
\usepackage{listings}
\usepackage{chngpage}
\setkeys{acs}{articletitle=true}
\usepackage{hyperref}
\hypersetup{
    colorlinks=true,
    linkcolor=black,
    filecolor=black,      
    urlcolor=black,
    citecolor=black,
}

\newcounter{cptDef}
\newcounter{cptProp}
\newcommand{\cptDef}{\refstepcounter{cptDef}\label{Defi \thecptDef} \roman{cptDef}\,.}
\newcommand{\cptProp}{\refstepcounter{cptProp}\label{Prop \thecptProp} \roman{cptProp}\,.}

\definecolor{codegreen} {rgb}{0.0 ,0.6 ,0.0 }
\definecolor{codered}   {rgb}{0.5 ,0.0 ,0.0 }
\definecolor{codegray}  {rgb}{0.5 ,0.5 ,0.5 }
\definecolor{codepurple}{rgb}{0.58,0.0 ,0.82}
\definecolor{backcolour}{rgb}{0.95,0.95,0.92}
\definecolor{mcol}{RGB}{ 0, 0, 135 }
\definecolor{mrep}{RGB}{46,95,0}

\newlist{mlist}{itemize}{1}
\setlist[mlist]{label=>,leftmargin=*,font=\normalsize}

\newcommand{\Definition}[1]{\noindent \textbf{Definition}\cptDef #1}
\newcommand{\Property}[1]{\noindent \textbf{Property}\cptProp #1}
\newcommand\commentaire[1]{}

\newcommand{\refCode}[1]{[Listing \ref{#1}]}
\newcommand{\refFig}[1]{[Figure \ref{#1}]}
\newcommand{\refTab}[1]{[Table \ref{#1}]}

\newcommand{\Tabxc}{\centering\arraybackslash}
\newcommand{\Tabxl}{\raggedleft\arraybackslash}
\newcommand{\Tabxr}{\raggedright\arraybackslash}
\newcolumntype{L}[1]{>{\raggedright\let\newline\\\arraybackslash\hspace{0pt}}m{#1}}
\newcolumntype{C}[1]{>{\centering\let\newline\\\arraybackslash\hspace{0pt}}m{#1}}
\newcolumntype{R}[1]{>{\raggedleft\let\newline\\\arraybackslash\hspace{0pt}}m{#1}}

\newcommand\TinkerHP{Tinker-\textit{HP}\,}
\newcommand\OpenACC{\textsc{OpenACC}\ }
\newcommand\CUDA{\textsc{CUDA}\ }
\newcommand\SIMT{\textsc{SIMT}}
\newcommand\FFt{\texttt{FFt}\ }
\newcommand\FFts{\texttt{FFt's}\ }
\newcommand{\cuFFt}{\texttt{cuFFt}\,}
\newcommand\gang{\texttt{gang}\ }
\newcommand\vecTor{\texttt{vector}\ }
\newcommand{\worker}{\texttt{worker}\ }
\newcommand{\GPU}{\emph{GPUs}\ }
\newcommand{\gpu}{\emph{GPU}\ }
\newcommand{\CPU}{\emph{CPUs}\ }
\newcommand{\cpu}{\emph{CPU}\ }
\newcommand{\MPI}{MPI\ }

\newcommand{\DP}{DP\ }
\newcommand{\FP}{FP\ }
\newcommand{\SP}{SP\ }
\newcommand\MP{MP\ }
\newcommand{\MD}{MD}

\newcommand{\dhfr}{\textsf{DHFR}\,}
\newcommand{\cox}{\textsf{COX}\,}
\newcommand{\protease}{\textsf{M\textsuperscript{pro}}\,}
\newcommand{\spike}{\textsf{SARS-Cov2 Spike-ACE2}\,}
\newcommand{\puddle}{\textsf{Puddle}\,}
\newcommand{\pond}{\textsf{Pond}\,}
\newcommand{\lake}{\textsf{Lake}\,}
\newcommand{\bay}{\textsf{Bay}\,}
\newcommand{\stmv}{\textsf{STMV}\,}
\newcommand{\sea}{\textsf{Sea}\, }

\newcommand{\numNeigMax}{\mathrm{Neig}_{max}}
\newcommand{\dcutoff}{\mathrm{d}_{\mathrm{cut}}}
\newcommand{\dbuffer}{\mathrm{d}_{\mathrm{buff}}}
\newcommand{\Block}{\mathrm{B}}
\newcommand{\Bcomp}{\mathrm{B}_{comp}}
\newcommand{\Bsize}{\Block_{size}}
\newcommand{\BOX}{\boldsymbol{\Omega}}
\newcommand{\Domain}{\boldsymbol{\psi}}
\newcommand{\DomainL}{\boldsymbol{\lambda}}
\newcommand{\Cell}{\boldsymbol{\omega}}
\newcommand{\NeigBlock}{\mathcal{B}}
\newcommand{\atom}{\alpha}
\newcommand{\nAtom}{\boldsymbol{n}}
\newcommand{\nAtomB}{\nAtom_b}
\newcommand{\nAtomP}{\nAtom_p}
\newcommand{\nAtomL}{\nAtom_l}

\newcommand{\Iblock}{\beta}
\newcommand{\funDi}{\mathsf{dist}}

\author{Olivier Adjoua}
\affiliation{Sorbonne Université, LCT, UMR 7616 CNRS, F-75005, Paris, France}
\author{Louis Lagardère*}
\affiliation{Sorbonne Université, LCT, UMR 7616 CNRS, F-75005, Paris, France}
\alsoaffiliation{Sorbonne Université, IP2CT, FR2622 CNRS, F-75005, Paris, France}
\author{Luc-Henri Jolly}
\affiliation{Sorbonne Université, IP2CT, FR2622 CNRS, F-75005, Paris, France}
\author{Arnaud Durocher}
\affiliation{Eolen, 37-39 Rue Boissière, 75116 Paris, France}
\author{Thibaut Very}
\affiliation{IDRIS, CNRS, Orsay, France}
\author{Isabelle Dupays}
\affiliation{IDRIS, CNRS, Orsay, France}
\author{Zhi Wang}
\affiliation{Department of Chemistry, Washington University in Saint Louis, USA}
\author{Théo Jaffrelot Inizan}
\affiliation{Sorbonne Université, LCT, UMR 7616 CNRS, F-75005, Paris, France}
\author{Fr\'{e}d\'{e}ric C\'{e}lerse}
\affiliation{Sorbonne Université, LCT, UMR 7616 CNRS, F-75005, Paris, France}
\alsoaffiliation{Sorbonne Université, CNRS, IPCM, Paris, France.}
\author{Pengyu Ren}
\affiliation{Department of Biomedical Engineering, The University of Texas at Austin, USA}
\author{Jay W. Ponder}
\affiliation{Department of Chemistry, Washington University in Saint Louis, USA}

\author{Jean-Philip Piquemal}
\affiliation{Sorbonne Université, LCT, UMR 7616 CNRS, F-75005, Paris, France}
\alsoaffiliation{Department of Biomedical Engineering, The University of Texas at Austin, USA}
\email{louis.lagardere@sorbonne-universite.fr,jean-philip.piquemal@sorbonne-universite.fr}

\title[\TinkerHP GPUs]
  {\TinkerHP : Accelerating Molecular Dynamics Simulations of Large Complex Systems with Advanced Point Dipole Polarizable Force Fields using GPUs and Multi-GPUs systems}

\abbreviations{IR,NMR,UV}

\begin{document}
\lstdefinestyle{codefortran}{
  language=[95]Fortran,                 
  backgroundcolor=\color{backcolour},   
  basicstyle=\ttfamily\scriptsize,      
  breakatwhitespace=false,              
  breaklines=false,                     
  captionpos=b,                         
  commentstyle=\color{codegreen},       
  deletekeywords={private,default},     
  extendedchars=true,                   
  keepspaces=true,                      
  keywordstyle=\color{blue},            
  keywordstyle=[2]\color{codered},      
  keywords=[2]{c\$acc,c\$acc\&,acc,parallel,loop,serial,atomic,update,gang,worker,vector,copy,copyin,copyout,present,create,async,attach,detach,default,private},    
  numbers=none,                         
  numbersep=5pt,                        
  numberstyle=\tiny\color{codegray},    
  rulecolor=\color{black},              
  showspaces=false,                     
  showstringspaces=false,               
  showtabs=false,                       
  stepnumber=1,                         
  stringstyle=\color{codepurple},       
  tabsize=3,                            
  title=\lstname                        
}

\lstset{style=codefortran}

%

\begin{abstract}
We present the extension of the \TinkerHP  package (Lagardère et al., Chem. Sci., 2018,9, 956-972) to the use of Graphics Processing Unit (GPU) cards to accelerate molecular dynamics simulations using polarizable many-body force fields. The new high-performance module allows for an efficient use of single- and multi-\GPU architectures ranging from research laboratories to modern supercomputer centers. After detailing an analysis of our general scalable strategy that relies on \OpenACC and \CUDA, we discuss the various capabilities of the package. Among them, the multi-precision possibilities of the code are discussed. If an efficient double precision implementation is provided to preserve the possibility of fast reference computations, we show that a lower precision arithmetic is preferred providing a similar accuracy for molecular dynamics while exhibiting superior performances. As \TinkerHP is mainly dedicated to accelerate simulations using new generation point dipole polarizable force field, we focus our study on the implementation of the AMOEBA model. Testing various NVIDIA platforms including 2080Ti, 3090, V100 and A100 cards, we provide illustrative  benchmarks of the code for single- and multi-cards simulations on large biosystems encompassing up to millions of atoms.
 The new code strongly reduces time to solution and offers the best performances to date obtained using the AMOEBA polarizable force field. Perspectives toward the strong-scaling performance of our multi-node massive parallelization strategy, unsupervised adaptive sampling and large scale applicability of the \TinkerHP code in biophysics are discussed. The present software has been released in phase advance on GitHub in link with the High Performance Computing community COVID-19 research efforts and is free for Academics (see https://github.com/TinkerTools/tinker-hp).
\end{abstract}

\section*{Introduction}
Molecular dynamics (MD) is a very active research field that is continuously progressing.\cite{DRORMDreview,DESRESreview} Among various evolutions, the definition of force fields themselves grows more complex. Indeed, beyond the popular pairwise additive models\cite{ponder2003force,CHARMMff,Amberff,OPLSFF,gromosff} that remain extensively used, polarizable force field (PFF) approaches are becoming increasingly mainstream and start to be more widely adopted\cite{reviewcompchem,annurev-biophys-070317-033349,melcr2019accurate,chemrevION}, mainly because accounting for polarizability is often crucial for complex applications and adding new physics to the model through the use of many-body potentials can lead to significant accuracy enhancements.\cite{melcr2019accurate} Numerous approaches are currently under development but a few methodologies such as the Drude \cite{Drudeprot,drudereview,Drude2020} or the AMOEBA \cite{ren2003polarizable,shi,DNAAMOEBA} models emerge. These models are more and more employed because of the alleviation of their main bottleneck: their larger computational cost compared to classical pairwise models. Indeed, the availability of High Performance Computing (HPC) implementations of such models within popular packages such as NAMD\cite{DrudeNAMD} or GROMACS \cite{DrudeGROMACS} for Drude or \TinkerHP \cite{Tinker-HP} for AMOEBA fosters the diffusion of these new generation techniques within the research community. This paper is dedicated to the evolution of the \TinkerHP package.\cite{Tinker-HP} The software, which is part of the Tinker distribution,\cite{TINKER8} was initially introduced as a double precision massively parallel MPI addition to Tinker dedicated to the acceleration of the various PFFs and non-polarizable force fields (n--PFFs) present within the Tinker package. The code was shown to be really efficient, being able to scale on up to tens of thousand cores on modern petascale supercomputers.\cite{Tinker-HP,jolly2019raising} Recently, it has been optimized on various platforms taking advantage of vectorization and of the evolution of the recent \CPU (Central Processing Units).\cite{jolly2019raising} However, in the last 15 years, the field has been increasingly using \GPU (Graphic Processor Unit) \cite{STONE2010116,routinems1,GROMACSGPUs} taking advantage of low precision arithmetic. Indeed, such platforms offer important computing capabilities at both low cost and high energy efficiency allowing for reaching routine microsecond simulations on standard \GPU cards with pair potentials.\cite{routinems1,routinems2}
Regarding the AMOEBA polarizable force field, the OpenMM package \cite{OpenMM7} was the first to propose an AMOEBA-GPU library that was extensively used within Tinker through the Tinker-OpenMM \gpu interface.\cite{TINKEROpenMM} The present contribution aims to address two goals: i) the design of an efficient native \TinkerHP GPU implementation; ii) the HPC optimization in a massively parallel context to address both the use of research laboratories clusters and modern multi-\GPU pre-exascale supercomputer systems. The paper is organized as follows. First, we will describe our \OpenACC port and its efficiency in double precision. After observing the limitations of this implementation regarding the use of single precision, we introduce a new \CUDA approach and detail the various parts of the code it concerns after a careful study of the precision. In both cases, we present benchmarks of the new code on illustrative large biosystems of increasing size on various NVIDIA platforms (including RTX 2080Ti, 3090, Tesla V100 and A100 cards). Then, we explore how to run on even larger systems and optimize memory management by making use of latest tools such as NVSHMEM \cite{potluri2016simplifying}. 

\section{\OpenACC Approach} \label{sec:OpenACC}
\subsection{Global overview and definitions} \label{subsec:global_overview}
\begin{figure}
    \centering
    \includegraphics[width=\linewidth]{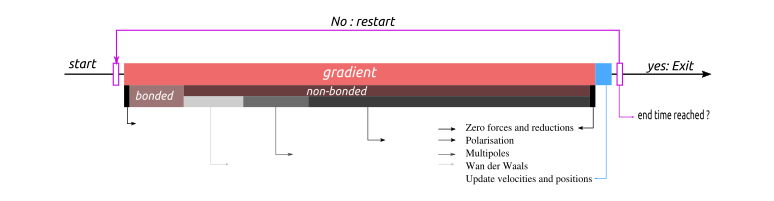}
    \caption{illustration of a MD timestep}
    \label{fig:Integrator}
\end{figure}

\TinkerHP is a molecular dynamics application with a MPI layer allowing a significant acceleration on \CPU. The core of the application is based on the resolution of the classical newton equations \cite{frenkel2001understanding,lagardere2018tinker} given an interaction potential (force field) between atoms. In practice, a molecular dynamic simulation consists into the repetition of the call to an integrator routine defining the changes of the positions and the velocities of all the atoms of the simulated system between two consecutive timesteps. The same process is repeated as many times as needed until the simulation duration is reached (see Figure \ref{fig:Integrator}). To distribute computations over the processes, a traditional three dimensional domain decomposition is performed on the simulation box ($\BOX$) which means that it is divided in subdomains ($\Domain$), each of which being associated to a MPI process. Then, within each timestep, positions of the atoms and forces are exchanged between processes before and after the computation of the forces. Additionally, small communications are required after the update of the positions to deal with the fact that an atom can change of subdomain during a timestep. This workflow is described in detail in reference\cite{lagardere2018tinker}.

\begin{figure}
    \centering
    \includegraphics[width=\linewidth]{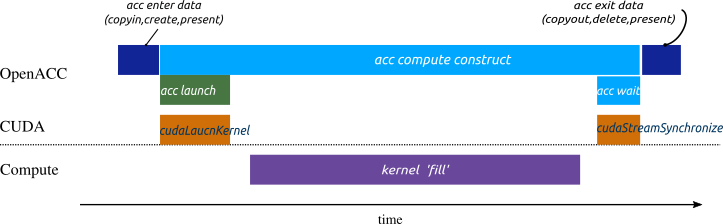}
    \caption{\OpenACC synchronous execution model on test kernel {\ttfamily <fill>} }
    \label{fig:OpenAcc_execution}
\end{figure}

\begin{figure}
    \centering
    \includegraphics[width=\linewidth]{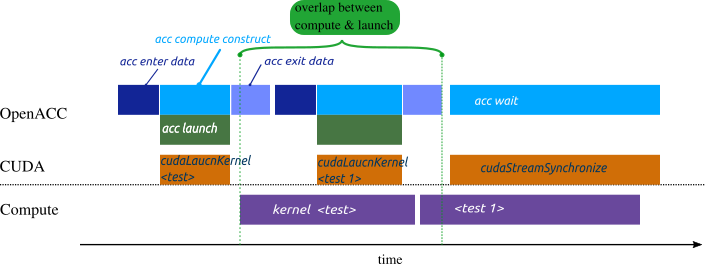}
    \caption{\OpenACC asynchronous execution on both kernels {\ttfamily <test> \& <test 1>}. }
    \label{fig:OpenAcc_async_execution}
\end{figure}

In recent years a new paradigm has emerged to facilitate computation and programming on \gpu devices.
In the rest of the text, we will denote as {\em kernel} the smallest piece of code made of instructions designed for a unique purpose. Thus, a succession of kernels might constitute a {\em routine} and a {\em program} can be seen as a collection of routines designed for a specific purpose. There are two types of kernels 
\begin{itemize}
    \item {\em Serial} kernels, mostly used for variable configuration
    \item {\em Loops} kernels, operating on multiple data sets
\end{itemize}
This programming style, named \OpenACC, \cite{10.1007/978-3-642-32820-6_85,10.5555/3175812} is a directive-based language similar to the multi-threading OpenMP paradigm with an additional complexity level. Since a target kernel is destined to be executed on \GPU, it becomes crucial to manage data between both \gpu and \cpu platforms. At the most elementary level, \OpenACC compiler interacts on a standard host (\cpu\!\!) kernel and generates a device (\gpu\!\!) kernel using directives implemented to describe its parallelism along with clauses to manage global data behaviour at both entry and exit point and/or kernel launch configuration \refFig{fig:OpenAcc_execution}.
This method offers two majors benefits. Unlike the low-level \CUDA programming language \cite{sanders2010cuda}, it takes only a few directives to generate a device kernel. Secondly, the same kernel is compatible with both platforms - \CPU and \GPU -. The portability along with all the associated benefits such as host debug is therefore ensured. However, there are some immediate drawbacks mainly because \CPU and \GPU do not share the same architecture, specifications and features. Individual \cpu cores benefit from a significant optimization for serial tasks, a high clock frequency and integrate vectorization instructions to increase processing speed. \GPU on the other hand were developed and optimized from the beginning for parallel tasks with numerous aggregations of low clock cores holding multiple threads. This means that it may be necessary to reshape kernels to fit device architecture in order to get appropriate acceleration.
Once we clearly exhibit a kernel parallelism and associate \OpenACC directives to offload it on device, it should perform almost as well as if it had been directly written in native \CUDA. Still, in addition to kernel launch instruction (performed by both \OpenACC and \CUDA) before the appropriate execution, there is a global data checking operation overhead which might slow down execution \refFig{fig:OpenAcc_execution}. However, it is possible to overlap this operation using asynchronous device streams in the kernel configuration \refFig{fig:OpenAcc_async_execution}. Under proper conditions and with directly parallel kernels, \OpenACC can already lead to an efficient acceleration close to the one reachable with \CUDA.


In the following, we will say that a kernel is \emph{semi-parallel} if one can find a partition inside the instructions sequence which does not share any dependency at all. A semi-parallel kernel is consequently defined \emph{parallel} if all instructions in the partition do not induce a race condition within its throughput.

Once a kernel is device compiled, its execution requires a configuration defining the associated resources provided by the device. With \OpenACC, these resources are respectively the total number of threads and the assignment stream. We can access the first one through the \emph{gang} and \emph{vector} clauses attached to a device compute region directive. A \gang is a collection of vectors inside of which every thread can share cache memory. All gangs run separately on device streaming multi-processors (SM) to process kernel instructions inside a stream where many other kernels are sequentially queued. \OpenACC offers an intermediate parallelism level between \gang and vector called \worker. This level can be seen as a gang subdivision. 
 
It is commonly known that \GPU are inefficient for sequential execution due to their latency. To cover up latency, each SM comes with a huge register file and cache memory in order to hold and run as many vectors as possible at the same time. Instructions from different gangs are therefore pipe-lined and injected in the compute unit \cite{Volkov:EECS-2016-143,sanders2010cuda}. From this emerges the kernel occupancy's concept which is defined as the ratio between the gang's number  concurrently running on one SM and the maximum gang number that can actually be held by this SM. 

\subsection{Global scheme}  \label{subsec:Global_scheme}
The \emph{parallel} computing power of \GPU is in constant evolution and the number of streaming multiprocessors (SM) is almost doubling with every generation. Considering their impressive compute potential in comparison to \CPU\!\!, one can assume that the only way to entirely benefit from this power is to offload the entire application on device. Any substantial part of the workflow of \TinkerHP should not be performed on the CPU platform. It will otherwise represent a bottleneck to performance in addition to requiring several data transfers. 
As for all MD applications, most of the computation lies in the evaluation of the forces. For the AMOEBA polarizable model, it takes around 97\% of a timestep to evaluate those forces when running sequentially on \cpu platform. Of these 97\%, around 10\% concern bonded forces and 90\% the non bonded ones, namely: polarization, (multipolar) permanent electrostatics and van der Waals. The polarization which includes the iterative resolution of induced dipoles largely dominates this part (see \refFig{fig:Integrator}). Non-bonded forces and polarization in particular will thus be our main focus regarding the porting and optimization. We will then benefit from the already present Tinker-HP MPI layer \cite{Tinker-HP,jolly2019raising} to operate on several \GPU\!\!. The communications can then be made directly between \GPU by using a \CUDA aware MPI implementation \cite{kraus2013introduction}.
The Smooth Particle Mesh Ewald method \cite{SPME,lagardere2015,lagardere2018tinker} is at the heart of both the permanent electrostatics and polarization non-bonded forces used in \TinkerHP, first through the iterative solution of the induced dipoles and then through the final force evaluation. It consists in separating the electrostatic energy in two independent pieces: real space and reciprocal space contributions. Let us describe our \OpenACC strategy regarding those two terms. 

\subsubsection{Real space scheme} \label{ssubsec:Real space}
Because the real space part of the total PME energy and forces has the same structure as \\ the van der Waals one, the associated \OpenACC strategy is the same. Evaluating real space energy and forces is made through the computation of pairwise interactions. Considering $n$ atoms, a total of $n(n-1)$ pairwise interactions need to be computed. This number is reduced by half because of the symmetry of the interactions. Besides, because we use a cutoff distance after which we neglect these interactions, we can reduce their number to being proportional to $n$ in homogeneous systems by using neighbor lists. The up-bound constant is naturally reduced to a maximum neighbors for every atoms noted as $\numNeigMax$.

\noindent The number of interactions is up-bounded by $n*\numNeigMax$. In terms of implementation, we have written the compute algorithm into a single loop kernel. As all the interactions are independent, the kernel is semi-parallel regarding each iteration. By making sure that energy and forces are added one at a time, the kernel becomes parallel. To do that, we can use atomic operations on \GPU which allow to make this operation in parallel and solve any race condition issue without substantially impacting parallel performance. By doing so, real space kernels looks like \refCode{code:real_space}

\begin{lstlisting}[caption={\OpenACC real space offload scheme.
    The kernel is offloaded on device using two of the three parallelism levels offered by \OpenACC. The first loop is broken down over gangs and gathers all data related to atom \texttt{iglob} using gang's shared memory through the private clause. \OpenACC vectors are responsible of the evaluation and the addition of forces and energy after resolving scaling factor if necessary. Regarding data management we make sure with the present clause that everything is available on device before the execution of the kernel.},
    label={code:real_space}
]
c$acc parallel loop gang default(present) async
c$acc&         private(scaling_data)
do i = 1,numLocalAtoms
   iglob = glob(i)  ! Get Atom i global id
   !Get Atom iglob parameter and positions
   ...
   !Gather Atoms iglob scaling interactions in 'scaling_data'
   ...
c$acc loop vector
   do k = 1, numNeig(i)
      kglob = glob( list(k,i) )
      ! Get Atom kglob parameter and positions
      ! Compute distance (d) between iglob & kglob
      if (d < dcut) then
         call resolve_scaling_factor(scaling_data)
         ...
         call Compute_interaction !inlined
         ...
         call Update_(energy,forces,virial)
      end if
   end do
end do
\end{lstlisting}

At first, our approach was designed to directly offload the \cpu vectorized real space compute kernels which use small arrays to compute pairwise interactions in hope to align memory access pattern at the vector level and therefore accelerate the code\cite{jolly2019raising}. This requires  each gang to privatize every temporary array and results in a significant overhead with memory reservation associated to a superior bound on the gang's number. Making interactions computation scalar helps us remove those constraints and double the kernel performance. The explanation behind this increase arises from the use of GPU scalar registers. Still, one has to resolve the scaling factors of every interactions. As it happens inside gang shared memory, the performance is slightly affected. However, we would benefit from a complete removal of this inside search. There are two potential drawbacks to this approach:
\begin{mlist}
    \item Scaling interactions between neighboring atoms of the same molecule can become very complex. This is particularly true with large proteins. Storage workspace can potentially affect shared memory and also kernel's occupancy.
    \item Depending on the interactions, there is more than one kind of scaling factor. For example, every AMOEBA polarization interactions need three different scaling factors. 
\end{mlist}
The best approach is then to compute scaling interactions separately in a second kernel. Because they only involve connected atoms, their number is small compared to the total number of non-bonded interactions. We first compute unscaled non-bonded interactions and then apply scaling correction in a second part. An additional issue is to make this approach compatible with the 3d domain decomposition. Our previous kernel then reads: \refCode{code:real_space2}.

\begin{lstlisting}[caption={ final \OpenACC real space offload scheme.
     This kernel is more balanced and exposes much more computational load over vectors. 'correct\_scaling' routine applies the correction of the scaling factors. This procedure appears to be much more suitable to device execution. },
     label={code:real_space2}
]
c$acc parallel loop gang default(present) async
do i = 1,numLocalAtoms
   iglob = glob(i)  ! Get Atom i global id
   !Get Atom iglob parameter and positions
   ...
c$acc loop vector
   do k = 1, numNeig(i)
      kglob = glob( list(k,i) )
      ! Get Atom kglob parameter and positions
      ! Compute distance (d) between iglob & kglob
      if (d < dcut) then
         call Compute_interaction !inlined
         ...
         call Update_(energy,forces,virial)
      end if
   end do
end do

call correct_scaling
\end{lstlisting}

\subsubsection{Reciprocal space scheme} \label{ssubsec:Reciprocal space}
\begin{figure}
    \centering
    \includegraphics[width=\textwidth]{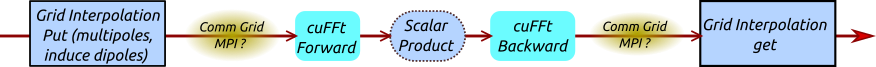}
    \caption{ Reciprocal space offload scheme. 
    Charge interpolation and Force interpolation are both written in a single kernel. They are naturally parallel except for the atomic contributions to the grid in the first one. The approach remains the same for data management between host and device as for real space: all data are by default device resident to prevent any sort of useless transfer. Regarding MPI communications, exchanges take place directly between \GPU through interconnection.} 
    \label{fig:reci_scheme}
\end{figure}

The calculation of Reciprocal space PME interactions essentially consists in five steps:
\begin{enumerate}
\item Interpolating the (multipolar) density of charge at stake on a 3D grid with flexible b-spline order (still, the implementation is optimized to use an order of 5 as it is the default and the one typically used with AMOEBA).
\item Switching to Fourier space by using a forward fast Fourier transform (\FFt\!\!)
\item Performing a trivial scalar product in reciprocal space
\item Performing a backward \FFt to switch back to real space
\item Performing a final multiplication by b-splines to interpolate the reciprocal forces
\end{enumerate}
Regarding parallelism, \TinkerHP uses a two dimensional decomposition of the associated 3d grid based on successive 1D \FFts. Here, we use the \cuFFt library.\cite{CUDAFFT}
 The \OpenACC offload scheme for reciprocal space is described in \refFig{fig:reci_scheme}

We just reviewed our offload strategy of the non-bonded forces kernels with \OpenACC, but the bonded ones remain to be treated. Also, the MPI layer has to be dealt with. The way bonded forces are computed is very similar to the real space ones, albeit simpler, which makes their offloading relatively straightforward. MPI layer kernels require, on the other hand, a slight rewriting as communications are made after a packing pre-treatment. In parallel, one does not control the throughput order of this packing operation. This is why it becomes necessary to also communicate the atom list of each process to their neighbors.
Now that we presented the main offload strategies, we can focus on some global optimizations regarding the implementation and execution for single and multi-\GPU.
Some of them lead to very different results depending on the device architecture. 

\subsection{Optimizations opportunities}  \label{subsec:Optimisations}

\commentaire{Plan et notes 
   Réarranger la nblist
   Précaluler le produit matrice vecteur
   Revisiter les opérations FFt
   Recouvrir les communications rec-rec}

\begin{mlist}
    \item A first optimization is to impose an optimal bound on the \vecTor size when computing pair interactions. In a typical setup, for symmetry reasons, the number of neighbors for real space interactions vary between zero and a few hundreds. Because of that second loop in \refCode{code:real_space2}, the smallest vector length ($32$) is appropriate to balance computation among the threads it contains. Another optimization concerns the construction of the neighbor lists. Let us remind that it consists in storing, for every atom, the neighbors that are closer than a cut distance ($\dcutoff$) plus a buffer $\dbuffer$. This buffer is related to the frequency at which the list has to be updated.
    To balance computation at the vector level and at the same time reduce warp discrepancy (as illustrated in \refFig{fig:warp_disc}) we have implemented a reordering kernel: we reorder the neighbor list for each atom so that the firsts are the ones under $\dcutoff$ distance.

    \begin{center}
        \includegraphics[width=0.95\textwidth]{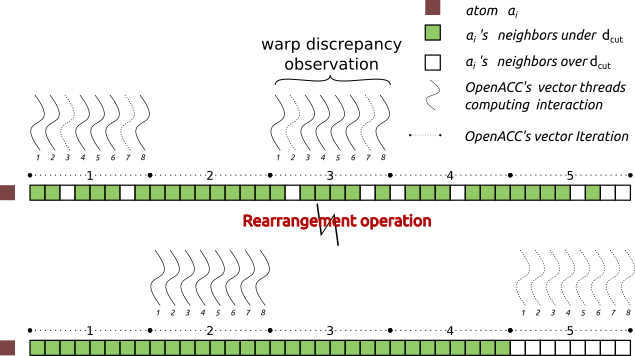}
        \captionsetup{type=figure}
        \caption{Illustration of compute balance due to the list reordering. Unbalanced computation in the first image induces an issue called warp discrepancy: a situation where all threads belonging to the same vector do not follow the same instructions. Minimizing that can increase kernel performance significantly since we ensure load balancing among each thread inside the vector.}
        \label{fig:warp_disc}
    \end{center}

    \item A second optimization concerns the iterative resolution of the induced dipoles. Among the algorithms presented in \cite{lipparini2014,lagardere2015}, the first method we offloaded is the preconditioned conjugated gradient (PCG). It involves a (polarization) matrix-vector product at each iteration. Here, the idea is to reduce the computation and favor coalesce memory access by precomputing and storing the elements of (the real space part) of the matrix before the iterations. As the matrix-vector product is being repeated, we see a performance gain starting from the second iteration. This improves performance but implies a memory cost that could be an issue on large systems or on \GPU with small memory capabilities. This overhead will be reduced at a high level of multi-device parallelism.

    \item An additional improvement concerns the two dimensional domain decomposition of the reciprocal space 3D grid involved with \FFt\!\!. The parallel scheme for \FFt used in \TinkerHP is the following for a forward transform: 
    
    { \itshape FFT(s) 1d dim(x) + x Transpose y + FFt(s) 1d dim(y) + y Transpose z + FFT(s) 1d dim(z) }.
    
    Each Transposition represents an all-to-all MPI communication which is the major bottleneck preventing most MD Applications using PME to scale across nodes \cite{PhillipsFFT,lagardere2018tinker,jolly2019raising}. Given the \GPU huge compute power, this communication is even more problematic in that context. On device, we use the \cuFFt \cite{CUDAFFT} library. Using many \cuFFt 1d batches is not as efficient as using less batches in a higher dimension. Indeed, devices are known to under perform with low saturation kernels.
    In order to reduce MPI exchanges and increase load on device, we adopted a simple 3d dimensional \cuFFt batch when running on a single device. On multiple \GPU, we use the following scheme based on a 1d domain decomposition along the z axis :
    \[ cuFFt(s)\; 2d\; dim(x,y) + y\; Transpose\; z + cuFFt(s)\; 1d\; dim(z) \]
    
    which gives a $25\%$ improvement compared to the initial approach.

    \item Profiling the application on a single device, we observed that real space computation is on average $5$ times slower than reciprocal space computation. This trend reverses using multiple \GPU because of the communications mentioned above. This motivated the assignment of these two parts in two different priority streams. Reciprocal space kernels along with MPI communications are queued inside the higher priority stream, and real space kernels - devoid of communications - can operate at the same time on the lower priority stream to recover communications. This is illustrated in \refFig{fig:async_recover}. 

    \begin{figure}
        \centering
        \includegraphics[width=\textwidth]{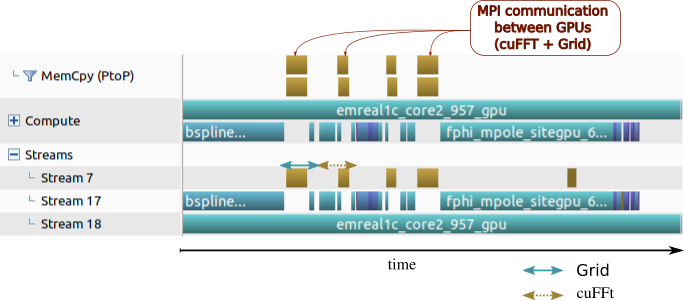}
        \caption{ Representation of \cuFFt\!'s communication/computation overlap using different streams for direct and reciprocal space. Real space computation kernels are assigned to asynchronous stream 18. reciprocal ones goes into high priority asynchronous stream 17. real space kernel therefore recovers \FFt grid exchanges. This profile was retrieved on 2\GPU. }
        \label{fig:async_recover}
    \end{figure}
\end{mlist}

\subsection{Simulation Validation and Benchmarks} \label{subsec:sim_val_res}

Here, we use the same bench systems as in reference \cite{Tinker-HP,jolly2019raising}: the solvated \dhfr protein, the solvated \cox protein and the \stmv virus, all with the AMOEBA force field, respectively made of \num{23558}, \num{171219} and \num{1066600} atoms. The molecular dynamics simulations were run in the NVT ensemble at $300K$ for \num{5} pico-seconds simulation using a (bonded/non-bonded) RESPA integrator with a 2fs outer timestep (and a 1fs inner timestep)\cite{tuckerman1992reversible} and the Bussi thermostat \cite{Bussi}. The performance was averaged over the complete runs. For validation purposes, we compared the potential energy, temperature and pressure of the systems during the first 50 timesteps with values obtained with \TinkerHP v1.2. Furthermore, these results were compared to Tinker-OpenMM in the same exact setup \cite{TINKEROpenMM}.

We can directly observe the technical superiority of the Quadro architecture compared to the Geforce one. Double precision (\DP\!\!) compute units of the V100 allow to vastly outperform the Geforce. In addition, by comparing the performance of the Geforce RTX to the one of the Quadro GV100, we see that Quadro devices are much less sensitive to warp discrepancy and non-coalesce data accessing pattern. It is almost as if the architecture of the V100 card overcomes traditional optimizations techniques related to parallel device implementation. However, we see that our pure \OpenACC implementation manages to deliver more performance than usual device MD application with PFF in \DP\!\!. The V100 results were obtained on the Jean-Zay HPE SGI 8600 cluster of the IDRIS supercomputer Center (GENCI-CNRS, Orsay, France) whose converged partitions are respectively made of 261 and 351 nodes. Each one is made of 2 Intel Cascade Lake 6248 processors (20 cores at 2,5 GHz) accelerated with 4 NVIDIA Tesla V100 SXM2 \GPU, interconnected through NVIDIA NVLink and coming respectively with \SI{32}{GB} of memory on the first partition and \SI{16}{GB} on the second. Here as in all the tests presented in this paper, all the MPI communications were made with a CUDA aware MPI implementation\cite{kraus2013introduction}. This results is very satisfactory as a single V100 card is at least ten times faster than an entire node of this supercomputer using only \CPU\!.

\begin{figure}[h]
    \centering
    \includegraphics[width=0.8\textwidth]{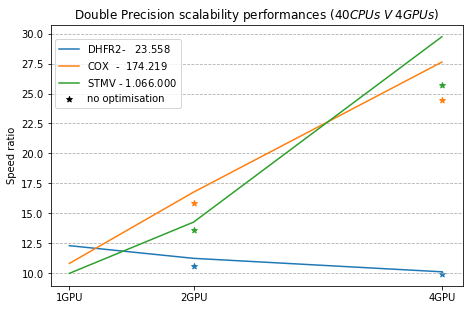}
    \caption{Performance ratio between single node \gpu and single node \cpu performance. Reference values can be found in \refTab{tab:speed_dbl}.}
    \label{fig:double_bench}
\end{figure}

Multi-device benchmark results compared with host-platform execution are presented in \refFig{fig:double_bench}. In practice, the \dhfr protein is too small to scale out. MPI communications overcomes the computations even with optimizations. On the other hand, \cox and \stmv systems show good multi-\GPU performances. 
Adding our latest MPI optimizations - \FFt reshaping and asynchronous compute between direct and reciprocal part of PME - allows a substantial gain in performances. 
We see that on a Jean-Zay node we can only benefit from the maximum communication bandwidth when running on the entire node, hence the relative inflexion point on the \stmv performances on \num{2} \GPU setup. Indeed, all devices are interconnected inside one node in such a way that they all share interconnection bandwidth. More precisely, running on \num{2} \GPU reduces bandwidth by four and therefore affects the scalabity. It is almost certain that results would get better on an interconnected node made exclusively of 2 \GPU. Those results are more than encouraging considering the fact that we manage to achieve them with a full \OpenACC implementation of \TinkerHP (direct portage of the reference \cpu code) in addition to some adjustments. 

\begin{table}
    \centering
    \begin{tabularx}{\textwidth}{| >{\Tabxc}X | >{\Tabxc}X | >{\Tabxc}X | >{\Tabxc}X | >{\Tabxc}X | >{\Tabxc}X | >{\Tabxc}X |} \hline
       {\small systems /devices (ns/day)} & {\small \CPU - one node} & {\small RTX 2080Ti } & {\small RTX 2080Ti + optim} & {\small V100} & {\small V100 + optim}  & {\small Tinker-OpenMM V100 } \vspace{2pt}\\ \hline
       \dhfr & 0.754  & 2.364 & 3.903 & 8.900 & 9.260 & 6.300 \\
       \cox  & 0.103  & 0.341 & 0.563 & 1.051 & 1.120 &     0.957 \\
       \stmv & 0.013  &  n/a  &   n/a & 0.111 & 0.126 & 0.130 \\ \hline
    \end{tabularx}
    \caption{Single device benchmark : MD production per day (ns/day). All simulations were run using a RESPA/\SI{2}{fs} setup. } 
    \label{tab:speed_dbl}
\end{table}

In summary, our \DP implementation is already satisfactory compared to other applications such as Tinker-OpenMM. Our next section concerns the porting of \TinkerHP in a downgraded precision.

\section{CUDA Approach} \label{sec:CUDA}
Even though we already have a robust \OpenACC implementation of \TinkerHP in double precision, the gain in terms of computational speed  when switching directly to single precision (\SP\!\!) is modest, as shown in \refTab{tab:sprec_openacc}, which is inconsistent with the \GPU computational capabilities.

This is more obvious for Geforce architecture devices since those cards do not possess \DP physical compute units and therefore emulate \DP Instructions.
According to \refTab{tab:devices_specs}, theoretical ratios of 2 and 31 are respectively expected from V100 and RTX-2080 Ti performances when switching from \DP to \SP which makes an efficient SP implementation mandatory.

\begin{table}[htbp]
    \centering
    \begin{tabularx}{0.75\textwidth}{| >{\Tabxc}X | >{\Tabxc}X  >{\Tabxc}X  >{\Tabxc}X |} \hline
                    & \dhfr & \cox  & \stmv \\ \hline
        V100        & 11.69 & 1.72 & 0.15 \\
        RTX-2080 Ti & 11.72 & 1.51 & n/a  \\ \hline
    \end{tabularx}
    \caption{Single precision MD production (\SI{}{ns/day}) within the \OpenACC implementation}
    \label{tab:sprec_openacc}
\end{table}

\begin{table}[htb]
    \centering
    \begin{tabularx}{\textwidth}{ L{4cm} | >{\Tabxl}X  >{\Tabxl}X  >{\Tabxl}X >{\Tabxl}X >{\Tabxl}X }
          \GPU              & \multicolumn{2}{c}{ Performances \SI{}{(Tflop/s)} } & memory & \multicolumn{2}{c}{Compute/Access} \\
                            & \DP         & \SP         &  bandwidth \SI{}{(GB/s)}  & \DP (Ops/8B) & \SP (Ops/4B) \\ \hline
        Quadro GV100        & \num{7.40}  & \num{14.80} & \num{870.0} & \num{68.04} & \num{ 68.04} \\
        Tesla V100 SXM2     & \num{7.80}  & \num{15.70} & \num{900.0} & \num{69.33} & \num{ 69.77} \\
        Geforce RTX-2080 Ti & \num{0.42}  & \num{13.45} & \num{616.0} & \num{ 5.45} & \num{ 87.33} \\
        Geforce RTX-3090    & \num{0.556} & \num{35.58} & \num{936.2} & \num{ 4.75} & \num{152.01} \\
    \end{tabularx}
    \caption{device hardware specifications}
    \label{tab:devices_specs}
\end{table}

\begin{figure}
    \centering
    \includegraphics[width=\textwidth]{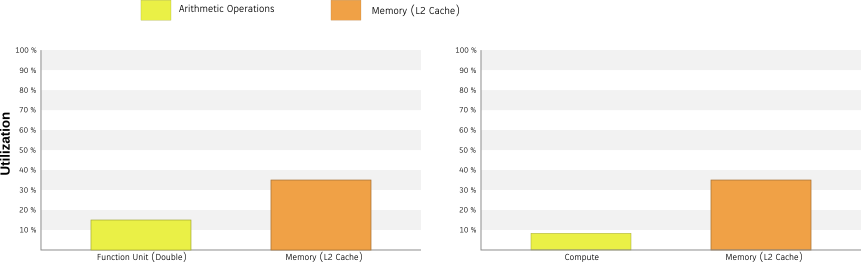}
    \caption{Profile of the matrix-vector compute kernel on the \dhfr system. The left picure is obtained with the double precision and the right one with simple precision. In both modes, results indicate an obvious latency issue coming from memory accessing pattern which prevents the device to reach its peak performance. }
    \label{fig:deep_profiling}
\end{figure}

In practice, instead of doubling the speed on V100 cards, we ended up noticing a \num{1.25} increase factor on V100 and \num{3} on RTX on \dhfr in \SP compared to \DP with the same setup. All tests where done under the assumption that our simulations are valid in this precision mode. More results are shown in \refTab{tab:sprec_openacc}.
Furthermore, a deep profile conducted on the kernels representing \TinkerHP's bottleneck - real space non-bonded interactions - in the current state reveals an insufficient exploitation of the GPU \SP compute power. \refFig{fig:deep_profiling} suggests that there is still room for improvements in order to take full advantage of the card's computing power and memory bandwidth both in \SP and \DP\!\!. In order to exploit device \SP computational power and get rid of the bottleneck exposed by \refFig{fig:deep_profiling}, it becomes necessary to reshape our implementation method and consider some technical aspects beyond \OpenACC\!\!'s scope.

\subsection{Global overview and definitions} \label{subsec:Global overview}
 As mentioned in the previous section, \GPU are most efficient with parallel computations and coalesce memory access patterns.
  The execution model combines and handles effectively two nested levels of parallelism. The high level concerns multithreading and the low level the SIMD execution model for vectorization \cite{zhou2002implementing,jolly2019raising}. This model stands for single instruction multiple threads (\SIMT) \cite{nickolls2010gpu}. When it comes to \gpu programming, \SIMT\ also includes control-flow instructions along with subroutine calls within SIMD level. This provides additional freedom of approach during implementation.
 To improve the results presented in the last paragraph \refTab{tab:sprec_openacc} and increase peak performance on computation and throughput, it is crucial to expose more computations in real space kernels and to minimize global memory accesses in order to benefit from cache \& shared memory accesses as well as registers. Considering \OpenACC paradigm limitations in terms of kernel description as well as the required low-level features, we decided to rewrite those specific kernels using the standard approach of low-level device programming in addition to \CUDA built-in intrinsics. In a following section, we will describe our corresponding strategy after a thorough review on precision.

\subsection{Precision study and Validation} \label{subsec:Precision study}

\Definition{ We shall call $\epsilon_p$ the machine precision (in \SP or \DP), the smallest floating point value such that $ 1+\epsilon_p > 1 $. They are respectively \num{1.2e-7} and \num{2.2e-16} in \SP and \DP.}

\Definition{Considering a positive floating point variable $a$, the machine precision $\epsilon_a$ attached to $a$ is  
\[ 1 + \epsilon > 1 \iff a + \epsilon_p*a > a \iff \epsilon_a = \epsilon_p*a \]
Therefore an error made for a floating point operation between $a$ and $b$ can be expressed as
\begin{equation}
 a \tilde{\oplus} b = ( a \oplus b )\;( 1 + \epsilon_p )
 \label{eq:num_prec}
\end{equation}
where $\tilde\oplus$ designates the numerical operation between $a$ and $b$. }

\Property{ Numerical error resulting from sequential reduction operations are linear while those resulting from parallel reduction are logarithmic. Thus, parallel reductions are entirely suitable to \GPU implementation as they benefit from both parallelism and accuracy.
\begin{figure}[h]
    \centering
    \includegraphics[width=\textwidth]{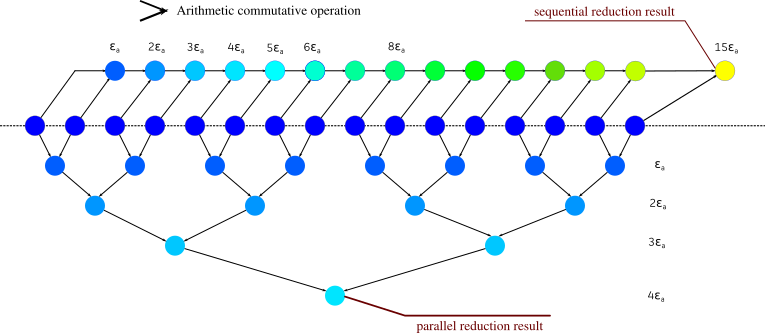}
    \caption{Illustration of the reduction operation on a 16 variables set. Each arithmetic operation generates an error $\epsilon_a$ that is accumulated during the sequential operation. On the other hand, parallel reduction uses intermediate variables to significantly reduce the error.}
    \label{fig:reduction_methods}
\end{figure}
}

Before looking further into the matter of downgrading precision, we have to make sure that Tinker-HP is able to work in this mode. Although it has been proven in the literature \cite{LEGRAND2013374,GROMACSGPUs,OpenMM7} that many MD applications are able to provide correct results with simple precision, extensive precision studies with polarizable force fields are lacking. 

When it comes to standard IEEE floating point arithmetic, regular \num{32} bits storage offers no more than \num{7} significant digits due to the mantissa. In comparison, we benefit from \num{16} significant digits with \DP \num{64} storage bits. Without any consideration on the floating number's sign, it is safe to assume that any application working with absolute values outside $[10^{-7}, 10^7]$ scope will fail to deliver sufficient accuracy when operating in complete \SP mode. This is called floating point overflow. To overcome this, the common solution is to use a mixed precision mode (\MP\!\!) which encompasses both standard \SP and a superior precision container to store variables subject to \SP overflowing.
In practice, most MD applications adopt \SP for computation and a higher precision for accumulation. Moreover, applications like Amber or OpenMM propose another accumulation method which rely on a different type of variable\cite{LEGRAND2013374}.

The description made in the previous section shows that energy and the virial evaluation are linear-dependent with the system's size. Depending on the complexity of the interaction in addition to the number of operations it requires, we can associate a constant error value $\epsilon_i$ to it. Thus, we can bound the error made on the computation of a potential energy with $N_{int}\epsilon_i < n\numNeigMax\epsilon_i$ where $N_{int}$ represents the number of interactions contributing to this energy and $\numNeigMax$ the maximum number of neighbor it involves per atom. As it is linear with respect to the system size we have to evaluate this entity with a \DP storage container. Furthermore, to reduce even more the accumulation of error due to large summation, we shall employ buffered parallel reduction instead of sequential one \refFig{fig:reduction_methods}.
On the other hand, we have to deal with the forces which remain the principal quantities which drive a MD simulation.
The error made for each atom on the non bonded forces is bound by $\numNeigMax*\epsilon_i$ depending on the cutoff.
However, each potential comes with a different $\epsilon_i$. In practice, the corresponding highest values are the one of both van der Waals and bonded potentials. The large number of pairwise interactions induced by the larger van der Waals cutoff in addition to the functional form which includes power of \num{14} (for AMOEBA) causes \SP overflowing for distances greater than \num{3}\AA. 
By reshaping the variable encompassing the pairwise distance, we get a result much closer to \DP since intermediate calculations do not overflow.  
Regarding the bonded potentials, $\epsilon_i^{bond}$ depends more on the conformation of the system.

Parameters involved in bond pairwise evaluation (spring stiffness,...) causes a \SP numerical error ($\epsilon_i^{bond}$) standing between \num{1e-3} and \num{1e-2} which frequently reach \num{1e-1} (following \eqref{eq:num_prec}) during summation process and this affects forces more than total energy. In order to minimize $\epsilon_i^{bond}$ we evaluate the distances in \DP before casting the result to \SP. In the end, $\epsilon_i^{bond}$ is reduced on the scope of [\num{1e-4},\num{1e-3}] which represents the smallest error we can expect from \SP\!. 

Furthermore, unlike the energy, a sequential reduction using atomic operations is applied to the forces. The resulting numerical error is therefore linear with the total number of summation operations. This is why we adopt a 64 bits container for those variables despite the fact they can be held in a 32 bits container.

Regarding the type of the 64 bits container, we analyze two different choices. First we have the immediate choice of a floating point. The classical mixed precision uses FP64 for accumulation and integration. Every MD applications running on GPU integrates this mode. It presents the advantage of being straightforward to implement. Second, we can use an integer container for accumulation: this is the concept of fixed point arithmetic introduced by Yates \cite{yates2009fixed}. To be able to hold a floating point inside an integer requires to define a certain number of bits to hold the decimal part. It is called the fixed point fractional bits. The left bits are dedicated to the integer part. Unlike the floating point, freezing the decimal position constrains the approximation precision but offers a correct accuracy in addition to deterministic operations. Considering a floating point value $x$ and an integer one $a$ and a fractional bits value ($\mathbf{fB}$), the relations establishing the transition back and forth between them - as $a=f(x)$ and $x=f^{-1}(a)$ - is defined as follow :
    \begin{align}
        a =& \;\mathtt{int}\left(\mathtt{round} \left(x\times 2^{\mathbf{fB}} \right) \right)  \\
        x =& \;\frac{\mathtt{real}(a)}{2^{\mathbf{fB}}}
    \end{align}
with $\mathtt{int}$ and $\mathtt{real}$ the converting functions and $\mathtt{round}$ the truncation function which extracts the integer part. When it comes to MD, fixed point arithmetic are an excellent tool: each \SP pairwise contribution is small enough to be efficiently captured by 64 bits fixed point. For instance, it takes only 27 bits to capture 8 digits after the decimal point with large place left for the integer part. For typical values observed with different systems size, we are far from the limit imposed by the integer part of the container. Inspired by the work of Walker, Götz et al. \cite{LEGRAND2013374}, we have implemented this feature inside \TinkerHP with the following configuration: a \num{34} fractional bits has been selected for forces accumulation, which leaves 30 bits for the integer part, thus setting the absolute limit value to $2^{29}$Kcal/mol$\cdot$\AA . For the energy, we only allocated \num{30} fractional bits given the fact that it grows linearly with the system size. 
Besides, using integer container for accumulation avoids dealing with \DP instructions which significantly affects performance on Geforce cards unlike Tesla ones. In summary, we should expect at least a performance or precision improvement from \FP\!.

\begin{figure}
    \centering
    \includegraphics[width=0.95\textwidth]{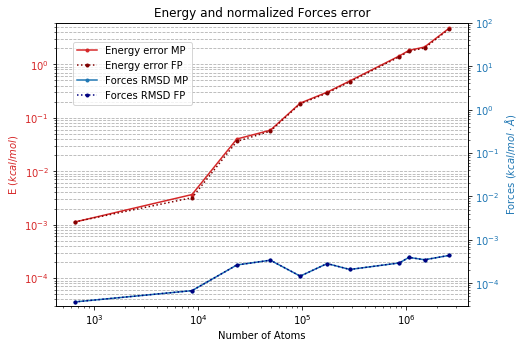}
    \caption{Absolute error between \DP implementation and both \FP and \MP implementations on total potential energy and Forces. Forces root mean square deviation between \DP and \MP for systems from \num{648} up to \num{2592000} atoms. As expected, both absolute errors in \SP and \FP are almost identical on the energy and they grow linearly with the system size. Logarithmic regression gives \num{0.99} value for the curve slope and up to \SI{5}{kcal/mol} for the largest system. However, the relative error for all systems is located under \num{7e-7} in comparison to \DP\!. One can also see that the error on the forces is independent of the system size the theory according to which, numerical errors generated from the computation do not depends of the system size.}
    \label{fig:error_study}
\end{figure}

A practical verification is shown in \refFig{fig:error_study}. 
In all cases, both \MP and \FP behave similarly. 
Forces being the driving components of \MD, the trajectories generated by our mixed precision implementation are accurate. However, as one can see, if errors remain very low for forces even for large systems, a larger error exists for energies, a phenomenon observed in all previous MD \GPU implementations.
Some specific post-treatment computations, like in a BAR free energy computation or NPT Simulations with a Monte-Carlo barostat, require accurate energies. In such a situation, one could use the \DP capabilities of the code for this post-processing step as \TinkerHP remains exceptionally efficient in \DP even for large systems. A further validation simulation in the NVE ensemble can be found in \refFig{fig:validation_prec} confirming the overall excellent stability of the code.

\subsection{Neighbor list} \label{subsec:Neighbor list}
\label{sec:Neighbor}

\begin{figure}
    \centering
    \includegraphics[scale=0.75]{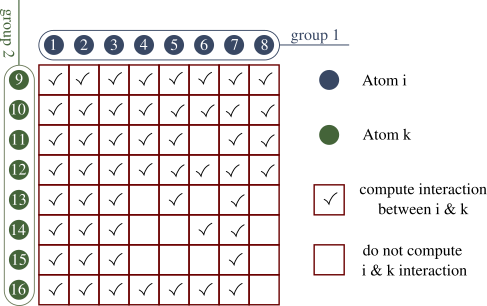}
    \caption{Representation of interactions between two groups of atoms within Tinker-HP. $\Bsize=8$ for the illustration}
    \label{fig:groups_inte}
\end{figure}

We want to expose the maximum of computation inside a kernel using the device shared memory. To do so, we consider the approach where a specific group of atoms interacts with another one in a block-matrix pattern (see \refFig{fig:groups_inte}). We need to load the parameters of the group of atoms and the output structures needed for computation directly inside cache memory and/or registers. On top of that, \CUDA built-in intrinsics can be used to read data from neighbor threads and if possible compute cross term interactions. Ideally, we can expose $\Bcomp=\Bsize^2$ computations without a single access to global memory, with $\Bsize$ representing the number of atoms within the group. With this approach, the kernel should reach its peak in terms of computational load.

A new approach of the neighbor list algorithm is necessary to follow the logic presented above. This method will be close to standard blocking techniques used in many MD applications \cite{GROMACSGPUs,OpenMM7}. Let us present the structure of the algorithm in a sequential and parallel - MPI - context.

\subsubsection{Box partitioning}

 Lets us recall that given a simulation box $\BOX$, a set of $\Cell_c$ with $c \in [0..\mathrm{N}c] $ forms a $\BOX$ partition if and only if
\[  \left\{ \begin{array}{c} \Cell_1 \cup \dotso \cup \Cell_{\mathrm{N}c} = \BOX  \\ \Cell_1 \cap \dotso \cap \Cell_{\mathrm{N}c} = \varnothing \end{array} \right .\]

We consider in the following that each group deals with interactions involving atoms within a region of space. In order to maximize $\Bcomp$ between every pair of groups, we must then ensure their spatial compactness. Moreover, all these regions need to define a partition of $\BOX$ to make sure we do not end up with duplicate interactions.
Following this reasoning, we might be tempted to group them into small spheres but it is impossible to partition a polygon with only spheres; not to mention the difficulties arising from the implementation point of view.

The MPI layer of \TinkerHP induces a first partition of $\BOX$ in $P$ subdomains $\Domain_p, p \in [0..P]$ where $P$ is the number of MPI processes. \TinkerHP uses the midpoint image convention \cite{midpoint} so that the interactions computed by the process assigned to $\Domain_p$ are the ones whose midpoint falls into $\Domain_p$. The approach used in \TinkerHP for the non bonded neighbor list uses a cubic partition $\Cell_c, c \in [1..{\mathrm{N}c}] $ of $\Domain_p$ and then collects the neighboring atoms among the neighboring cells of $\Cell_c$. 
Here, we proceed exactly in the same way with two additional conditions to the partitioning. First, the number of atoms inside each cell $\Cell_c$ must be less or equal than $\Bsize$. Second, we must preserve a common global numbering of the cells across all domains $\Domain_p$ to benefit from a unique partitioning of $\BOX$.

\noindent Once the first partitioning in cells is done, an additional sorting operation is initiated to define groups so that each of them contains exactly $\Bsize$ spatially aligned atoms following the cell numbering (note that because of the first constrain mention earlier, one cell can contain atoms belonging to a maximum of two groups). More precisely, the numbering of the cells follows a one dimensional representation of the three dimension of the simulation box.

Now, we want to find the best partitioning of $\Domain_p$ in groups that will ensure enough proximity between atoms inside a group, minimizing the number of neighboring groups and consequently maximizing $\Bcomp$.

\begin{figure}
    \centering
    \includegraphics[width=\textwidth]{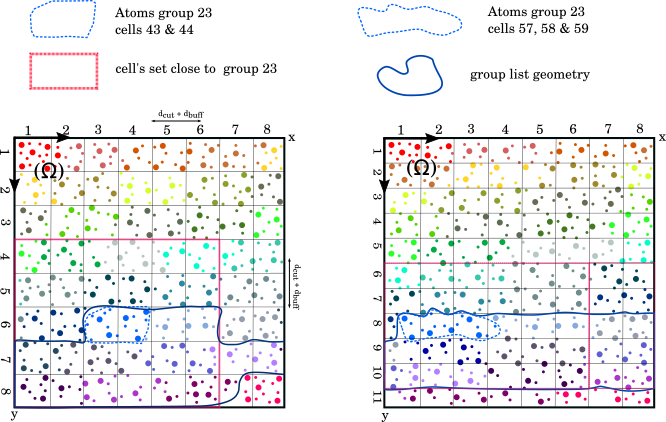}
    \caption{ Illustration of a two dimensional partition along with groups for a box of water. The left Figure shows a 64 cells partition of $\BOX$ while the right one refines this partitioning into 88 cells. The groups are defined by re-indexing the atoms following the cells numbering and their maximum size.
    Here $\Bsize=16$. A unique color is associated to every atom belonging to the same group. No cell contains more than $\Bsize$ atoms or 2 groups. Once a group is selected -23-, searching for its neighboring groups is made through the set of cells (with respect to the periodic boundary conditions) near to the cells it contains - 43 \& 44 -. Once this set is acquired, all the group indexes greater or equal than the selected group constitute the actual list of neighbors to take the symmetry of the interactions into account. We see that the group's shape modulates the group neighborhood as illustrated with the right illustration and a spatially flat group 23. }
    \label{fig:BoxPartition}
\end{figure}

When the partitioning generates too flat domains, each group might end up having too many neighboring groups. The optimal cell shape (close to a sphere) is the cube but we must not forget the first constrain and end up with a very thin partition either. However, atom groups are not affected by a partition along the inner most contiguous dimension in the cell numbering. We can exploit this to get better partitioning. \refFig{fig:BoxPartition} illustrates and explains the scheme on a two dimensional box. Partitioning is done in an iterative manner by cycling on every dimension. We progressively increase the number of cells along each dimension starting on the contiguous one until the first condition is fulfilled. During a parallel run, we keep track of the cell with the smallest number of atoms with a reduction operation. This allows to have a global partitioning of $\BOX$ and not just $\Domain_p$.

Now that we do dispose of a spatial rearrangement of the atoms into groups, we need to construct pair-lists of all interacting groups according to the cutoff distance plus an additional buffer to avoid reconstructing it at each timestep.

Groups are built in such a way that it is straightforward to jump from groups indexing to cells indexing. We chose to use an adjacency matrix which is \GPU suitable and compatible with MPI parallelism.

Once it is built, the adjacency matrix directly gives the pair-list. Regarding the storage size involved with this approach, note that we only require single bit to tag pair-group interactions. This results in an $ \lceil \frac{\nAtomL}{\Bsize} \rceil^2 $ bits occupation which equals to $ \lceil \frac{n}{\Bsize} \rceil^2 \frac{1}{8} $ bytes. $\nAtomL$ represents the number of atoms which participates to real space evaluation on a process domain $(\Domain_p)$. 
Of course, in terms of memory we cannot afford a quadratic reservation. However the scaling factor $ \lceil \frac{1}{\Bsize} \rceil^2 \frac{1}{8} $ is small enough even for the smallest value of $\Bsize$ set to 32 corresponding to device warp size. Not to mention that, in the context of multi-device simulation, the memory distribution is also quadratic. The pseudo-kernel is presented in \refCode{code:adjMat}

\begin{lstlisting}[caption=
{Adjacency matrix construction pseudo-kernel. We browse through all the cells and for each one we loop on their neighbors. It is easy to compute their ids since we know their length as well as their arrangement. Given the fact that all cells form a partition of the box, we can apply the symmetrical condition on pair-cells and retrieve the groups inside thanks to the partitioning condition, which ensures that each cell contains at most two groups.},
                   label={code:adjMat} ]
c$acc parallel loop default(present)
do i = 1, numCells
   celli = i
   !get blocks_i inside celli
   ...
c$acc loop vector
   do j = 1, numCellsNeigh
      ! Get cellj with number
      ...
      ! Get blocks_j inside cellk
      ...
      ! Apply symmetrical condition
      if ( cellj > celli ) cycle
c$acc loop seq
      do bi in blocks_i
         do bj in blocks_j
c$acc atomic
            set matrix(bj,bi) to 1
         end do
      end do
   end do
end do
\end{lstlisting}

Once the adjacency matrix is built, a simple post-processing gives us the adjacency list with optimal memory size and we can use the new list on real space computation kernels following the process described in the introduction of this subsection and illustrated in \refFig{fig:groups_inte}. In addition, we benefit from a coalesced memory access pattern while loading blocks data and parameters when they are spatially reordered.

\subsubsection{List filtering} \label{ssubsec:list filtering}

\begin{figure}
    \centering
    \includegraphics[scale=1]{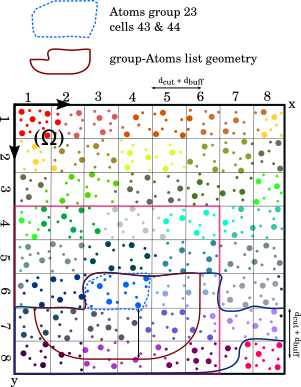}
    \caption{ Starting from the situation illustrated by Figure \ref{fig:BoxPartition}, we represent the geometry resulting from the filtering process. We significantly reduce group 23 neighborhood with the list filtering- it decreases from 144 atoms with the first list to 77 with the filtered one -.  $\Bcomp$ increases which corresponds to more interactions computed within each group pair. }
    \label{fig:group-Atoms}
\end{figure}

It is possible to improve the performance of the group-group pairing with a similar approach to the list reordering method mentioned in the \OpenACC optimizations section above. By filtering every neighboring group, we can get a list of atoms which really belong to a group's neighborhood. The process is achieved by following the rule:
    \[ \atom \in \NeigBlock_I \quad \text{if} \;  \exists \atom_i \in \Iblock_I  \quad \text{such that } \;  \funDi(\atom_i, \atom) \leq \dcutoff + \dbuffer . \]
$\atom$ and $\atom_i$ are atoms, $\Iblock$ represents a group of $\Bsize$ atoms, $\NeigBlock$ is the neighborhood of a group and $\funDi : (\mathbb{R}^3 \times \mathbb{R}^3 ) \to \mathbb{R}$ \; is the euclidean distance. \newline
An illustration of the results using the filtering process is depicted in \refFig{fig:group-Atoms}. 

\begin{figure}
    \centering
    \includegraphics[width=\textwidth]{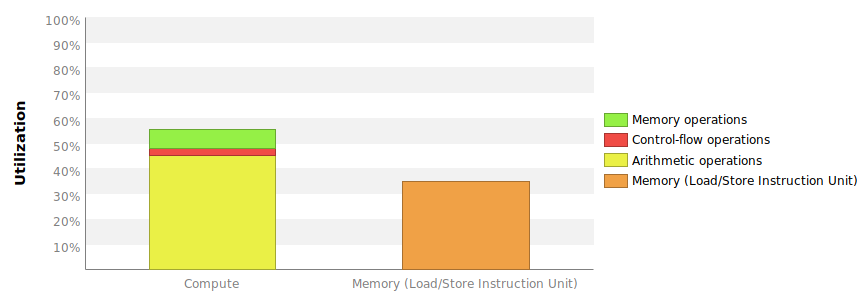}
    \caption{ Real space kernel profiling results in mixed precision using our new group-Atoms list. }
    \label{fig:prof_mix2}
\end{figure}

When the number of neighbor atoms is not a multiple of $\Bsize$, we create phantom atoms to complete the actual neighbor lists.
A drawback of the filtering process is a loss of coalesced memory access pattern. As it has been entirely constructed in parallel, we don't have control of the output order. Nonetheless, this is compensated by an increase of $\Bcomp$ for each interaction between groups, as represented by \refFig{fig:group-Atoms}. In practice, we measure a \(75\%\) performance gain between the original list and the filtered one for the van der Waals interaction kernel. Moreover, \refFig{fig:prof_mix2} (deep profile of the previous bottleneck kernel: matrix-vector product ) shows a much better utilisation of the device computational capability. We apply the same strategy for the other real space kernels (electrostatics and polarization).

\subsection{PME separation} \label{subsec:PME separation}
As mentioned above, the Particle Mesh Ewald method separates electrostatics computation in two - real and reciprocal space -. A new profiling of \TinkerHP in single-device mixed precision mode with the latest developments shows that the reciprocal part is the new bottleneck. More precisely, real space performs \num{20}\% faster than reciprocal space within a standard PME setup. Moreover, reciprocal space is even more a bottleneck in parallel because of the additional MPI communications induced by the \cuFFt Transformations. This significantly narrows our chances of benefiting from the optimizations mentioned in the previous optimization subsection.
However, as both parts are independent, we can distribute them on different MPI processes in order to reduce or even suppress communications inside \FFts. During this operation, a subset of \GPU are assigned to reciprocal space computation only. Depending on the system size and the load balancing between real and reciprocal spaces, we can break through the scalability limit and gain additional performance on a multi-device configuration.

\subsection{Mixed precision validation} \label{ssubsec:Mixed precision validation}
To validate the precision study made above,
we compare a 1 nanosecond long simulation in \DP on CPU (\TinkerHP 1.2) in a constant energy setup (NVE) with the exact same run using both GPU \MP \& \FP implementations.

\begin{figure}
    \includegraphics[width=0.95\textwidth]{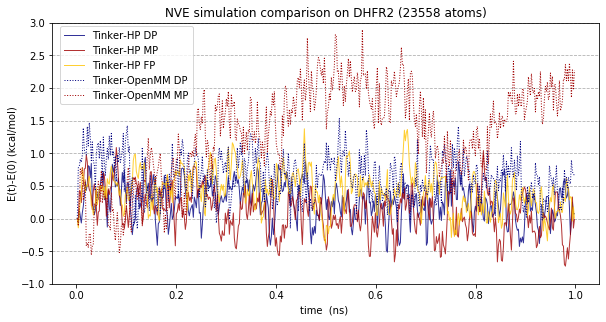}
    \caption{ Variation of the total energy during a NVE molecular dynamics simulation of the \dhfr protein in \DP and \MP and \FP. Energy fluctuations are respectively within \SI{1.45}{Kcal/mol} \SI{1.82}{Kcal/mol} \SI{1.75}{Kcal/mol} \SI{1.69}{Kcal/mol} \SI{3.45}{Kcal/mol} from \TinkerHP \DP \SP \FP and Tinker-OpenMM \DP \MP\!.}
    \label{fig:validation_prec}
\end{figure}

We used the solvated \dhfr protein and the standard velocity verlet integrator with a 0.5fs timestep, 12\AA\ and 7\AA\ cutoff distances respectively for van der Waals and real space electrostatics and a convergence criteria of \num{1e-6} for the polarization solver. A grid of $64\times 64\times 64$ was used for reciprocal space with fifth order splines. We also compare our results with a trajectory obtained with Tinker-OpenMM in \MP in the exact same setup, see \refFig{fig:validation_prec}.

The energy is remarkably conserved along the trajectories obtained with \TinkerHP in all cases: using \DP\!\!, \MP or \FP with less oscillations than with Tinker-OpenMM with \MP\!.

\subsection{Available Features}

The main features of \TinkerHP have been offloaded to GPU such as its various integrators like the multi-timestep integrators: RESPA1 and BAOAB-RESPA1 \cite{pushing} which allow up to a \SI{10}{fs} timestep with PFF (this required to create new neighbor lists to perform short-range non-bonded interactions computations for both van der Waals and electrostatics). Aside from Langevin integrators, we ported the Bussi\cite{Bussi} (which is the default) and the Berendsen thermostats, as well as the Monte-Carlo and the Berendsen barostats. 
We also ported free energy methods such as the Steered Molecular Dynamics \cite{Celerse2019} and van der Waals soft cores for alchemical transformations, as well as the enhanced sampling method Gaussian Accelerated Molecular Dynamics\cite{miao2015gaussian}.
Even if it is not the main goal of our implementation as well optimized software suited to such simulations exist, we also ported the routines necessary to use standard non-polarizable force fields such as CHARMM\cite{CHARMMff}, Amber\cite{Amberff} or OPLS\cite{jorgensen1996development}. Still, we obtained already satisfactory performances with these models despite a simple portage, the associated numbers can be found in supporting information and further optimization is ongoing. 
On top of all these features that concern a molecular dynamics simulation, we ported the "analyze" and "minimize" program of Tinker-HP, allowing to run single point calculations as well as geometry optimizations. All these capabilities are summed up in \refTab{tab:tinker_features}

\begin{table}[htb]
    \centering
    \begin{tabularx}{\textwidth}{l| >{\Tabxr}X }
        Programs      & dynamic; analyze; minimize \\
        Integrator    & VERLET(default); RESPA; RESPA1; BAOAB-RESPA; BAOAB-RESPA1 \\
        Force fields  & AMOEBA; CHARMM/AMBER/OPLS \\
        Miscellaneous & Steered MD (SMD); Gaussian Accelerated MD; Restrains Groups; Soft Cores; Plumed \\
        Thermostat    & Bussi(default); Berendsen \\
        Barostat      & Berendsen(default); Monte Carlo \\
    \end{tabularx}
    \caption{Available features in the initial \TinkerHP GPU release}
    \label{tab:tinker_features}
\end{table}

\subsection{Performance and Scalability Results}
 \label{subsec:Performances and scalability results}

We ran benchmarks with various systems on a set of different \GPU in addition to Tesla V100 nodes of the Jean-Zay supercomputer. We also ran the whole set of tests on the Irène Joliot Curie ATOS Sequana supercomputer V100 partition to ensure for the portability of the code. We used two different integrators: ( \SI{2}{fs} RESPA along with \SI{10}{fs} BAOAB-RESPA1 with heavy hydrogens).
For each system, we performed \SI{2.5}{ps} and \SI{25}{ps} MD simulations with RESPA and BAOAB-RESPA1 respectively and average the performance on the complete runs. Van der Waals and real space electrostatics cutoffs were respectively set to \num{9} and \num{7}\AA\, plus \num{0.7}\AA\,  neighbor list buffer for RESPA, \num{1}\AA\ for BAOAB-RESPA1. We use the Bussi thermostat with the RESPA integrator. Induced dipoles were converged up to a \num{1e-5} convergence threshold with the conjugate gradient solver and a diagonal preconditioner\cite{lagardere2015}. The test cases are: water boxes within the range of \num{96000} atoms (i.e. \puddle) up to \num{2592000} atoms (i.e \bay), the \dhfr, \cox and the Main Protease of Sars-Cov2 proteins (\protease) \cite{JaffrelotInizan2020} as well as the \stmv virus. Table \ref{tab:perfs_rtx} gathers all single devices performances and \refFig{fig:perfs_V100} illustrates the multi-device performance.

On a single GPU, the BAOAB-RESPA1 integrator performs almost twice as fast as RESPA in all cases - \num{22.53} to \SI{42.83}{ns/day} on \dhfr, \num{0.57} to \SI{1.11}{ns/day} for the \stmv virus. Regarding the RESPA integrator, results compared with those obtained in \DP \refTab{tab:speed_dbl} are now consistent with the Quadro V100 theoretical performance \refTab{tab:speed_dbl}. Moreover, we observe a significant improvement on single V100 cards with \DP in comparison to the \OpenACC implementation which shows that the algorithm is better suited to the architecture. However, this new algorithm considerably underperforms on Geforce architecture.
For instance, for the cox system the speed goes from \SI{0.65}{ns/j} with the \OpenACC implementation to \SI{0.19}{ns/j} with the adapted CUDA implementation on Geforce RTX-2080 Ti. 
This is obviously related to architecture constrains (lack of \DP Compute units, sensitivity to SIMD divergence branch, instruction latency) and shows that there is still room for optimization. \TinkerHP is tuned to select the quickest algorithm depending on the target device. Concerning \MP performance on Geforce Cards, we finally get the expected ratio compared with \DP: increasing computation per access improves the use of the device \refTab{tab:devices_specs}. Geforce RTX-2080 Ti and GV100 results are close  until the \cox test case which is consistent with their computing power, but GV100 performs better for larger systems. It is certainly due to the difference in memory bandwidth which allows GV100 to perform better on memory bound kernels and to reach peak performance more easily. For example, most of PME reciprocal space kernels are memory bounded due to numerous accesses to the three dimensional grid during the building and extracting process. 

A further comparison between architectures is given in supporting information.

For \FP simulations, as expected, we don't see any performance difference with \MP on V100 cards unlike Geforce ones which exhibit an 8\% acceleration in average as the \DP accumulation is being replaced by an integer one ( an instruction natively handled by compute cores ).
 \refTab{tab:perfs_omm} shows the performance of Tinker-OpenMM: with the same RESPA framework, \TinkerHP performs 12 to 30\% better on GV100 when the system size grows. With Geforce RTX-2080 Ti the difference is slightly more steady except for the \lake test case: around 18\% and 25\% better performance with \TinkerHP respectively with \MP and \FP compared to Tinker-OpenMM.

The parallel scalability starts to be effective above \num{100000} atoms. This is partly because of the mandatory host synchronisations needed by MPI and because of the difference in performance between synchronous and asynchronous computation under that scale (for example \dhfr production drops to \SI{12}{ns/day} when running synchronous with the host). Kernel launching times are almost equivalent to their execution time and they do not overlap.
Each \gpu on the Jean Zay Supercomputer comes with a \SI{300}{GB/s} interconnection NVlink bandwidth. 4\GPU per node, all of them being interconnected, represents then a \SI{100}{Gb/s} interconnection for each \gpu pairs. The third generation PCI-Express bridge to the host memory only delivers \SI{16}{Gb/s}. 
With the RESPA integrator operating on a full node made of 4 Tesla V100, the speed ratio grows from \num{1.14} to \num{1.95} respectively from \puddle to \bay test cases  in comparison to a single device execution. The relatively balanced load between pme real and reciprocal space allows to break through the scalability limit on almost every run with 2 \GPU with PME separation enabled. Performance is always worse on 4\GPU with 1\gpu dedicated to the reciprocal space and the others to the direct space for the same reason mentioned earlier (direct/reciprocal space load balancing). We also diminished the communication overhead by overlapping communication and computation.
Note that on a complete node of Jean-Zay with 4 \GPU, the bandwidth is statically shared between all of them, which means that the performance showed here on two \GPU is less that what can be expected on a node that would only consist in two \GPU interconnected through NVlink.
With the BAOAB-RESPA1 integrator, ratios between a full node and a single device vary from \num{1.07} to a maximum of \num{1.58}. Because of the additional short range real space interactions, it is unsuited for pme separation,
yet the reduced amount of FFt offers a potential for scalability higher than RESPA. 
Such a delay in the strong scalability is understandable given the device computational speed, the size of the messages size imposed by the parallel distribution and the configuration run. The overhead of the MPI layer for \stmv with BAOAB-RESPA1 and 4\GPU bench is on average 41\% of a timestep. It consists mostly in FFt grid exchange in addition to the communication of dipoles in the polarization solver. This is an indication of the theoretical gain we can obtain with an improvement of the interconnect technology or the MPI layer. Ideally, we can expect to produce \SI{2.63}{ns/day} on a single node instead of \SI{1.55}{ns/day}. It is already satisfactory to be able to scale on such huge systems and further efforts will be made to improve multi-\GPU results in the future.

\begin{center}[ht]
    \includegraphics[width=\textwidth]{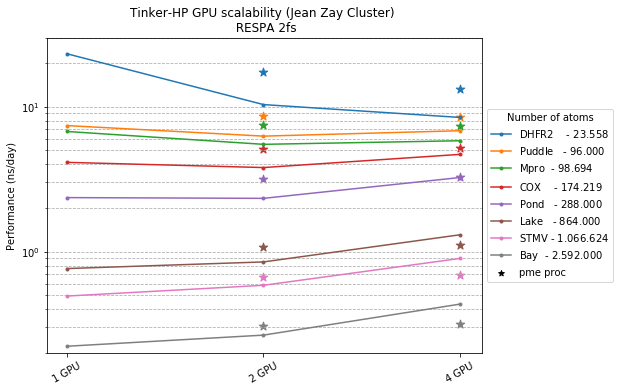}
    \includegraphics[width=\textwidth]{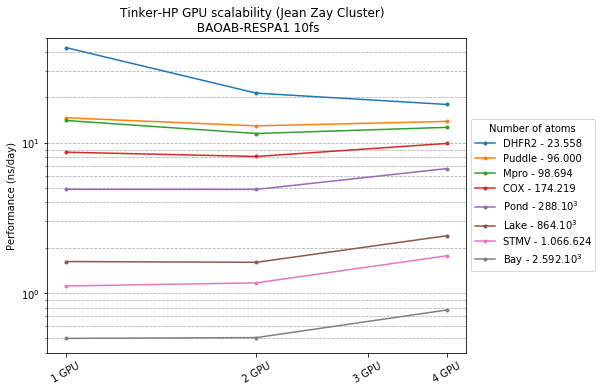}
    \captionsetup{type=figure}
    \caption{Single node mixed precision scalability on the Jean Zay Cluster (V100) using the AMOEBA polarizable force field}
    \label{fig:perfs_V100}
\end{center}

\begin{table}[htbp]
    \centering
    \begin{tabularx}{\textwidth}{ L{3cm} | >{\Tabxc}X >{\Tabxc}X >{\Tabxc}X >{\Tabxc}X >{\Tabxc}X >{\Tabxc}X >{\Tabxc}X }
        Systems                    & \dhfr & \protease & \cox & \pond & \lake & \stmv & \bay \\ \hline \hline
                                   & \multicolumn{7}{|c}{ DP Quadro GV100  } \\ \hline
        {\small RESPA 2fs }        & 11.24 &     2.91  & 1.76 & 1.08  & 0.36  & 0.24  & 0.11 \\
   {\scriptsize BAOAB-RESPA1 10fs} & 22.03 &     6.09  & 3.61 & 2.25  & 0.76  & 0.53  & 0.24 \\ \hline
                                   & \multicolumn{7}{|c}{ MP } \\ \hline
        {\small RESPA 2fs }        & 21.75 &     5.98  & 3.69 & 2.20  & 0.70  & 0.44  & 0.20 \\
   {\scriptsize BAOAB-RESPA1 10fs} & 40.73 &    12.80  & 3.61 & 4.58  & 1.49  & 1.01  & 0.46 \\ \hline
                                   & \multicolumn{7}{|c}{ FP } \\ \hline
        {\small RESPA 2fs }        & 21.46 &     5.82  & 3.57 & 2.12  & 0.67  & 0.43  & 0.20 \\
   {\scriptsize BAOAB-RESPA1 10fs} & 40.65 &    12.65  & 7.77 & 4.52  & 1.47  & 1.00  & 0.45 \\ \hline
                                   & \multicolumn{7}{|c}{ MP Geforce RTX-2080 Ti } \\ \hline
        {\small RESPA 2fs }        & 22.52 &     5.35  & 3.21 & 1.82  & 0.54  & 0.33  & 0.15 \\
   {\scriptsize BAOAB-RESPA1 10fs} & 43.81 &     11.85 & 7.06 & 4.06  & 1.24  & 0.82  & n/a  \\ \hline
                                   & \multicolumn{7}{|c}{ FP } \\ \hline
        {\small RESPA 2fs }        & 24.95 &      5.73 & 3.45 & 1.95  & 0.57  & 0.35  & 0.16 \\
   {\scriptsize BAOAB-RESPA1 10fs} & 47.31 &     12.78 & 7.63 & 4.35  & 1.32  & 0.87  & n/a  \\ \hline
                                   & \multicolumn{7}{|c}{ MP Geforce RTX-3090 }  \\ \hline
        {\small RESPA 2fs }        & 29.14 &      7.79 & 4.76 & 2.81  & 0.91  & 0.60  & 0.28 \\
   {\scriptsize BAOAB-RESPA1 10fs} & 52.80 &     15.79 & 9.61 & 5.52  & 1.81  & 1.23  & 0.59 \\ \hline
                                   & \multicolumn{7}{|c}{ FP } \\ \hline
        {\small RESPA 2fs }        & 32.00 &      8.37 & 5.10 & 3.02  & 0.96  & 0.64  & 0.30 \\
   {\scriptsize BAOAB-RESPA1 10fs} & 57.67 &     17.20 &10.46 & 5.96  & 1.90  & 1.32  & 0.63 \\ \hline
    \end{tabularx}
    \caption{\TinkerHP performances in (ns/day) on different devices and precision modes}
    \label{tab:perfs_rtx}
\end{table}

\begin{table}[htbp]
    \centering
    \begin{tabularx}{
    \textwidth}{ L{3cm} | >{\Tabxc}X >{\Tabxc}X >{\Tabxc}X >{\Tabxc}X >{\Tabxc}X >{\Tabxc}X >{\Tabxc}X }
        Systems         & \dhfr & \protease & \cox & \pond & \lake & \stmv & \bay \\ \hline \hline
        Quadro GV100    & 17.53 &      4.50 & 2.56 &  1.68 &  0.56 &  0.34 &  n/a \\
        Geforce RTX-2080 Ti & 18.97 &  4.37 & 2.63 &  1.66 &  0.55 &  0.28 &  n/a \\
    \end{tabularx}
    \caption{ Tinker-OpenMM Mixed precision performances assessed with RESPA framework}
    \label{tab:perfs_omm}
\end{table}

\section{ Towards larger systems }

As one of the goals of the development of \TinkerHP is to be able to treat (very) large biological systems such as protein complexes or entire viruses encompassing up to several millions of atoms (as it is already the case with the \CPU implementation \cite{Tinker-HP,jolly2019raising} by using thousands of \CPU cores), we 

review in the following section the scalability limit of the GPU implementation in terms of system size knowing that \GPU don't have the same memory capabilities

: where classical \cpu nodes routinely benefit from more than \SI{128}{GB} of memory, the most advanced Ampere \gpu architecture holds up to \SI{40}{GB} of memory.  

\subsection{ \TinkerHP Memory management model }
MD with 3D spatial decomposition has its own pattern when it comes to memory distribution among \MPI processes. We use the midpoint rule to compute real space interactions as it is done in the CPU implementation. 

In practice, it means that each process holds information about its neighbors (to be able to compute the proper forces). More precisely, a domain $\Domain_q$ belongs to the neighborhood of $\Domain_p$ if the minimum distance between them is under some cutoff distance plus a buffer. To simplify data exchange between processes, we transfer all positions in a single message, the same thing is done with the forces.

An additional filtering is then performed to list the atoms actually involved in the interactions computed by a domain  $\Domain_p$.
An atom - $\atom \in \BOX$ - belong to domain $\Domain_p$ 's interaction area ($\DomainL_p$) if the distance between this atom and the domain is below $\frac{\dcutoff + \dbuffer}{2}$.%

\begin{figure}
\includegraphics[scale=1]{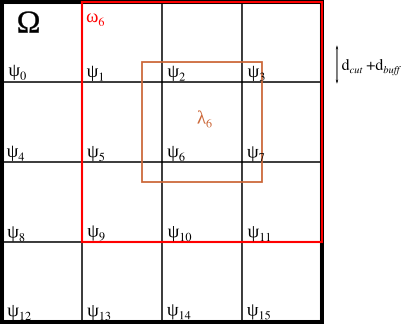}
\caption{Two dimensional spatial decomposition of a simulation box with \MPI distribution across 16 processes. $\omega_6$ collects all the neighboring domains of $\Domain_6$. Here $\nAtomB < \nAtom $ }
\label{fig:tinker_space_distribution}
\end{figure}

Lets us call $\nAtomP$ the number of atoms belonging to $\Domain_p$, $\nAtom_b$ the number of atoms belonging to a process domain and its neighbors and $\nAtomL$ the number of atoms inside $\DomainL_p$. This is illustrated in \refFig{fig:tinker_space_distribution}.

One can see that all data reserved with a size proportional to $\nAtomP$ are equally distributed among processes. Those with size proportional to $\nAtomB$ are only partially distributed. This means that these data structures are not distributed if all domains $\Domain_p$ are neighbors. This is why in practice the distribution only takes place at that level with a relatively high number of process - more than 26 at least on a large box with 3d domain decomposition -. On the other hand, data allocated with a size proportional to $\nAtomL$ (like the neighbor list) are always more distributed when the number of processes increases.

On top of that, some data remain undistributed (proportional to $\nAtom$) like the atomic parameters of each potential energy term. Splitting those among MPI processes would severely increase the communication cost which we can not afford. As we cannot predict how one atom will interact and move inside $\BOX$, the best strategy regarding such data is to make it available to each process. Reference \TinkerHP reduces the associated memory footprint by using MPI shared memory space: only one parameter data instance is shared among all processes within the same node.

No physical shared memory exist between \GPU of a node and

the only way to deal with undistributed data is by replicating them on each device which is quickly impractical for large systems.

In the next section, we detail a strategy allowing to circumvent this limitation.

\subsection{NVSHMEM feature implementation}
As explained above, distribution of parameter data would necessarily result in additional communications. Regarding data exchange optimizations between \gpu devices, NVIDIA develops a new library based on the OpenSHMEM \cite{chapman2010introducing} programming pattern which is called NVSHMEM \cite{potluri2016simplifying}. This library provides thread communication routines which operate on a symmetric memory on each device meaning that it is possible to initiate device communication inside kernels and not outside with an API like MPI. The immediate benefit of such approach resides in the fact that communications are automatically recovered by kernels instructions and can thereby participate to recover device internal latency.
This library allows to distribute $\nAtom$ scale data over devices within one node.

Our implementation follows this scheme: divide a data structure (an array for instance) across devices belonging to the same node following the global numbering of the atoms and access this data inside a kernel with the help of NVSHMEM library. To do that, we rely on a NVHSMEM feature which consists in storing a symmetric memory allocation address in a pointer on which arithmetic operations can be done. Then, depending on the address returned by the pointer, either a global memory access (GBA) or a remote memory access (RMA) is instructed to fetch the data. The implementation requires a Fortran interface to be operational since NVSHMEM source code is written in the C language. Moreover, an additional operation is required for every allocation performed by the NVSHMEM specific allocator to make the data allocated accessible through \OpenACC kernels.

\begin{figure}
    \centering
    \includegraphics{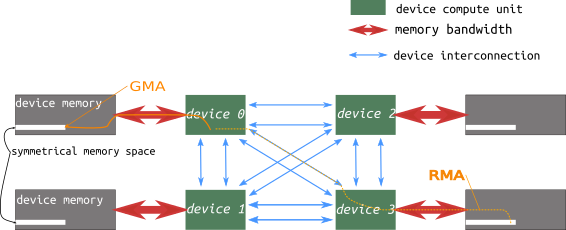}
    \caption{NVSHMEM memory distribution pattern across a four interconnected devices node. A symmetric reserved space is allocated by the NVSHMEM library at initialization. Thus, data is equally split across all devices in order. Every time a device needs to access data allocated with NVSHMEM, either a GMA or RMA is issued. }
    \label{fig:nvshmem_dist}
\end{figure}

Such a singular approach affects performances since additional communications have to be made inside kernels. Furthermore, all communications do not follow a special pattern that would leave room for optimizations, meaning that each device accesses data randomly from the others depending on the atoms involved in the interactions it needs to compute. In order to limit performance loss, we can decide which data is going to be split across devices and which kernels are going to be involved with this approach. In practice, we use this scheme for the parameters of the bonded potential.

Doing so, we distribute most of the parameter data - (torsions, angles, bonds,...) - and therefore reduce the duplicated memory footprint.

\subsection{ Perspectives and additional results}

During our NVSHMEM implementation, we were able to detect and optimize several memory wells. For instance, the adjacency matrix described in section \ref{sec:Neighbor}. has a quadratic memory requirement following the groups of atoms. This means that this represents a potential risk of memory saturation on a single device. To prevent this, we implemented a buffer limit on this matrix to construct the pair-group list piece by piece. We also implemented algorithms that prioritize computing and searching over storing where ever needed - essentially scaling factor reconstruction -. In the end, \TinkerHP is able to reach a performance of \SI{0.15}{ns/day} for a \num{7776000} atoms water box with the AMOEBA force field and the BAOAB-RESPA1 integrator on a single V100 and scale-out to \SI{0.25}{ns/day} on a complete node of Jean Zay on the same system.

\begin{figure}[H]
    \centering
    \includegraphics[width=\textwidth]{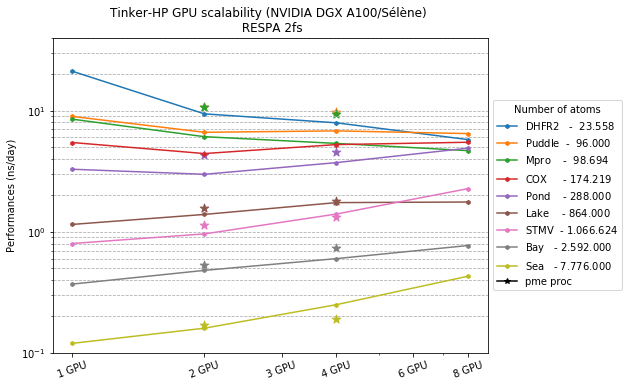}
    \includegraphics[width=\textwidth]{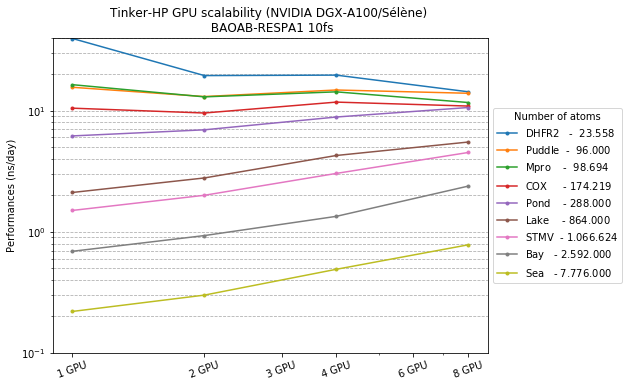}
    \caption{Performance and one node scalibility results with the AMOEBA force field}
    \label{fig:scalability_A100}
\end{figure}

We also had the opportunity to test our implementation on the latest generation NVIDIA \gpu Ampere architecture: the Selene supercomputer which is made of nodes consisting in DGX-A100 servers. A DGX-A100 server contains eight A100 graphic cards with 40 GB of memory each and with latest generation inter-connection NVIDIA Switches. The results we obtained on such a node with the same systems as above in the same RESPA and BAOAB-RESPA1 framework are listed in \refTab{tab:perfs_clusters} and \refFig{fig:scalability_A100}.

We observe an average of \num{50}\% of performance gain for systems larger than \num{100000} atoms on a single A100 compared to a single V100 card. Also, the more efficient interconnection between cards (NV-switch compared to NV-link) allows to scale better on several \GPU with the best performances ever obtained with our code on all the benchmark systems, the larger ones making use of all the 8 cards of the node.
Although the code is designed to do so, the latency and the speed of the inter-node interconnection on the present Jean-Zay and Selene supercomputers did not allow us to scale efficiently across nodes, even on the largest systems.
Jean Zay provides \SI{32}{GB/s} of network interconnection between nodes so that each \gpu pair has access to a \SI{16}{GB/s} bandwidth. Unlike the \SI{100}{GB/s} shared between each \gpu inside a node, we expect inter-node transit times to be \num{6.25} to \num{12.5} time slower without taking the latency into account. This is illustrated by the experiment summarized in \refTab{tab:multi_node_sea} as we observe the sudden increase of the overhead of the MPI layer relative to the total duration of a timestep when running on two nodes. In this case, changing the domain decomposition dimension to limit the number of neighbouring process quadruples the production and exposes the latency issue (expressed here by the difference between the fastest and the slowest MPI process). In a multi-node context the bottleneck clearly lies in the inter-node communications.  
The very fast evolution of the compilers, as well as the incoming availability of new classes of large pre-exascale supercomputers may improve this situation in the future. Presently, the use of multiple nodes for a single trajectory is the subject of active work within our group and results will be shared in due course. Still, one can already make use of several nodes with the present implementation by using methods such as unsupervised adaptive sampling as we recently proposed \cite{JaffrelotInizan2020}. Such pleasingly parallel approach already offers the possibility to use hundreds (if not thousands!) of \gpu cards simultaneously.

\begin{table}[ht]
\begin{tabularx}{0.95\textwidth}{| L{5cm} | >{\Tabxc}X >{\Tabxc}X | >{\Tabxc}X >{\Tabxc}X |} \hline
        \# GPU                    &   \multicolumn{2}{c|}{4}  & \multicolumn{2}{c|}{8} \\
        Domain Decomposition      &    3d    &  1d     &   3d    &   1d     \\ \hline
        Production speed (ns/day) &   0.252  &  0.265  &   0.02  &   0.08   \\
        MPI Layer (\%)            &  24      &  22     &  97     &  91      \\
        MPI Latency (\%)          &   1      &   1     &   4     &  11      \\ \hline
\end{tabularx}
\caption{multi-node performance on Jean Zay  with \sea system and BAOAB-RESPA1. Here the latency designates the time difference between the fastest and the slowest process.}
\label{tab:multi_node_sea}
\end{table}
\begin{table}[H]
\begin{tabularx}{0.95\textwidth}{| L{3cm}  R{2.cm} | >{\Tabxc}X >{\Tabxc}X | >{\Tabxc}X >{\Tabxc}X |} \hline
        \multirow{2}{1.7cm}{\small Systems}&  \multirow{2}{2cm}{{\small Size \newline (NumAtoms) }} & \multicolumn{2}{c|}{ Jean Zay (V100)}  & \multicolumn{2}{c|}{ Selene (A100) } \\ [1ex]
       & & {\small Perf (\SI{}{ns/day}) \newline - 1GPU } & {\small Best Perf (\SI{}{ns/day}) \newline - \#GPU} & {\small Perf (\SI{}{ns/day}) \newline - 1GPU } & {\small Best Perf (\SI{}{ns/day}) \newline - \#GPU} \\ \hline \hline
       \dhfr     & \num{  23558} & \num{43.83} & \num{43.83}      & \num{44.96} & \num{44.96}       \\
       \puddle   & \num{  96000} & \num{14.63} & \num{15.76} \;-4 & \num{15.57} & \num{17.57}       \\
       \protease & \num{  98694} & \num{14.03} & \num{14.57} \;-4 & \num{16.36} & \num{17.47} \;-4  \\
       \cox      & \num{ 174219} & \num{08.64} & \num{10.15} \;-4 & \num{10.47} & \num{11.75} \;-4  \\
       \pond     & \num{ 288000} & \num{04.90} & \num{06.72} \;-4 & \num{06.18} & \num{10.60} \;-8  \\
       \lake     & \num{ 864000} & \num{01.62} & \num{02.40} \;-4 & \num{02.11} & \num{05.50} \;-8  \\
       \stmv     & \num{1066624} & \num{01.11} & \num{01.77} \;-4 & \num{01.50} & \num{04.51} \;-8  \\
       \spike    & \num{1509506} & \num{00.89} & \num{01.55} \;-4 & \num{01.32} & \num{04.16} \;-8  \\
       \bay      & \num{2592000} & \num{00.50} & \num{00.77} \;-4 & \num{00.59} & \num{02.38} \;-8  \\
       \sea      & \num{7776000} & \num{00.15} & \num{00.25} \;-4 & \num{00.22} & \num{00.78} \;-8  \\ \hline
\end{tabularx}
\caption{Performance synthesis and scalability results on the Jean Zay (V100) and Selene (A100) machines. MD production in ns/day with the AMOEBA polarizable force field}
\label{tab:perfs_clusters}
\end{table}

\section*{Conclusion}
We presented the native \TinkerHP  multi-\GPU - multi-precision acceleration platform. The new code is shown to be accurate and scalable across multiple \GPU cards offering unprecedented performances and new capabilities to deal with long timescale simulations on large realistic systems using polarizable force fields such as AMOEBA. The approach strongly reduces the time to solution offering to achieve routine simulations that would have require thousands of \CPU on a single \gpu card. Overall, the \gpu-accelerated \TinkerHP reaches the best performances ever obtained for AMOEBA simulations and extends the applicability of polarizable force fields. The package is shown to be compatible with various computer \gpu system architectures ranging from research laboratories to modern supercomputers.

Future work will focus on adding new features (sampling methods, integrators ...) and on further optimizing the performance on multi-nodes /multi-\GPU to address the exascale challenge. We will improve the non-polarizable force field simulations capabilities as we will provide the high performance implementations of additional new generation polarizable many-body force fields models such as AMOEBA+ \cite{CW1, CW2}, SIBFA\cite{SIBFA} and others. We will continue to develop the recently introduced adaptive sampling computing strategy enabling the simultaneous use of hundreds (thousands) of \gpu cards to further reduce time to solution and deeper explore conformational spaces at high-resolution.\cite{JaffrelotInizan2020}  With such exascale-ready simulation setup, computations that would have taken years can now be achieved in days thanks to \GPU. Beyond this native \TinkerHP GPU platform and its various capabilities, an interface to the Plumed library \cite{bonomi2019promoting} providing additional methodologies for enhanced-sampling, free-energy calculations and the analysis of molecular dynamics simulations is also available. Finally, the present work, that extensively exploits low precision arithmetic, highlights the key fact that high-performance computing (HPC) grounded applications such as \TinkerHP can now efficiently use converged GPU-accelerated supercomputers, combining HPC and artificial intelligence (AI) such as the Jean Zay machine to actually enhance their performances.

\section*{Acknowledgements}
This work was made possible thanks to funding from the European Research Council (ERC) under the European Union's Horizon 2020 research and innovation programme (grant agreement No 810367), project EMC2. This project was initiated in 2019 with a "Contrat de Progrès" grant from GENCI (France) in collaboration with HPE and NVIDIA to port Tinker-HP on the Jean Zay HPE SGI 8600 GPUs system (IDRIS supercomputer center, GENCI-CNRS, Orsay, France) using \OpenACC.  FC acknowledges funding from the French state funds managed by the CalSimLab LABEX and the ANR within the Investissements d’Avenir program (reference ANR11-IDEX-0004-02) and support from the Direction Génerale de l’Armement (DGA) Maîtrise NRBC of the French Ministry of Defense. Computations have been performed at GENCI on the Jean Zay machine (IDRIS) on grant no A0070707671 and on the Irène Joliot Curie ATOS Sequana X1000 supercomputer (TGCC, Bruyères le Chatel, CEA, France) thanks to PRACE COVID-19 special allocation (projet COVID-HP). We thank NVIDIA (Romuald Josien and François Courteille, NVIDIA France) for offering us access to A100 supercomputer systems (DGX-A100 and Selene DGX-A100 SuperPod machines). PR and JWP are grateful for support by National Institutes of Health (R01GM106137 and R01GM114237).

\section*{Code availability}
The present code has been released in phase advance in link with the High Performance Computing community COVID-19 research efforts.  The software is freely accessible to Academics via GitHub : https://github.com/TinkerTools/tinker-hp

\section*{Competing interest}
The authors declare no competing interests.

\section*{Additional information}

Additional informations regarding the performance using non polarizable force fields as well as a comparison between peak performance reachable by Tinker-HP in term of FLOP/s can be find in SI.
This information is available free of charge via the Internet at \url{http://pubs.acs.org}

\newpage
\listoffigures
\newpage
\listoftables

\newpage
\bibliography{sample}

\providecommand{\latin}[1]{#1}
\makeatletter
\providecommand{\doi}
  {\begingroup\let\do\@makeother\dospecials
  \catcode`\{=1 \catcode`\}=2 \doi@aux}
\providecommand{\doi@aux}[1]{\endgroup\texttt{#1}}
\makeatother
\providecommand*\mcitethebibliography{\thebibliography}
\csname @ifundefined\endcsname{endmcitethebibliography}
  {\let\endmcitethebibliography\endthebibliography}{}
\begin{mcitethebibliography}{59}
\providecommand*\natexlab[1]{#1}
\providecommand*\mciteSetBstSublistMode[1]{}
\providecommand*\mciteSetBstMaxWidthForm[2]{}
\providecommand*\mciteBstWouldAddEndPuncttrue
  {\def\EndOfBibitem{\unskip.}}
\providecommand*\mciteBstWouldAddEndPunctfalse
  {\let\EndOfBibitem\relax}
\providecommand*\mciteSetBstMidEndSepPunct[3]{}
\providecommand*\mciteSetBstSublistLabelBeginEnd[3]{}
\providecommand*\EndOfBibitem{}
\mciteSetBstSublistMode{f}
\mciteSetBstMaxWidthForm{subitem}{(\alph{mcitesubitemcount})}
\mciteSetBstSublistLabelBeginEnd
  {\mcitemaxwidthsubitemform\space}
  {\relax}
  {\relax}

\bibitem[Hollingsworth and Dror(2018)Hollingsworth, and Dror]{DRORMDreview}
Hollingsworth,~S.~A.; Dror,~R.~O. Molecular Dynamics Simulation for All.
  \emph{Neuron} \textbf{2018}, \emph{99}, 1129 -- 1143\relax
\mciteBstWouldAddEndPuncttrue
\mciteSetBstMidEndSepPunct{\mcitedefaultmidpunct}
{\mcitedefaultendpunct}{\mcitedefaultseppunct}\relax
\EndOfBibitem
\bibitem[Dror \latin{et~al.}(2012)Dror, Dirks, Grossman, Xu, and
  Shaw]{DESRESreview}
Dror,~R.~O.; Dirks,~R.~M.; Grossman,~J.; Xu,~H.; Shaw,~D.~E. Biomolecular
  Simulation: A Computational Microscope for Molecular Biology. \emph{Annual
  Review of Biophysics} \textbf{2012}, \emph{41}, 429--452, PMID:
  22577825\relax
\mciteBstWouldAddEndPuncttrue
\mciteSetBstMidEndSepPunct{\mcitedefaultmidpunct}
{\mcitedefaultendpunct}{\mcitedefaultseppunct}\relax
\EndOfBibitem
\bibitem[Ponder and Case(2003)Ponder, and Case]{ponder2003force}
Ponder,~J.~W.; Case,~D.~A. \emph{Advances in protein chemistry}; Elsevier,
  2003; Vol.~66; pp 27--85\relax
\mciteBstWouldAddEndPuncttrue
\mciteSetBstMidEndSepPunct{\mcitedefaultmidpunct}
{\mcitedefaultendpunct}{\mcitedefaultseppunct}\relax
\EndOfBibitem
\bibitem[Huang \latin{et~al.}(2017)Huang, Rauscher, Nawrocki, Ran, Feig,
  de~Groot, Grubm{\"u}ller, and MacKerell]{CHARMMff}
Huang,~J.; Rauscher,~S.; Nawrocki,~G.; Ran,~T.; Feig,~M.; de~Groot,~B.~L.;
  Grubm{\"u}ller,~H.; MacKerell,~A.~D. CHARMM36m: an improved force field for
  folded and intrinsically disordered proteins. \emph{Nature methods}
  \textbf{2017}, \emph{14}, 71--73\relax
\mciteBstWouldAddEndPuncttrue
\mciteSetBstMidEndSepPunct{\mcitedefaultmidpunct}
{\mcitedefaultendpunct}{\mcitedefaultseppunct}\relax
\EndOfBibitem
\bibitem[Maier \latin{et~al.}(2015)Maier, Martinez, Kasavajhala, Wickstrom,
  Hauser, and Simmerling]{Amberff}
Maier,~J.~A.; Martinez,~C.; Kasavajhala,~K.; Wickstrom,~L.; Hauser,~K.~E.;
  Simmerling,~C. ff14SB: Improving the Accuracy of Protein Side Chain and
  Backbone Parameters from ff99SB. \emph{Journal of Chemical Theory and
  Computation} \textbf{2015}, \emph{11}, 3696--3713, PMID: 26574453\relax
\mciteBstWouldAddEndPuncttrue
\mciteSetBstMidEndSepPunct{\mcitedefaultmidpunct}
{\mcitedefaultendpunct}{\mcitedefaultseppunct}\relax
\EndOfBibitem
\bibitem[Jorgensen \latin{et~al.}(1996)Jorgensen, Maxwell, and
  Tirado-Rives]{OPLSFF}
Jorgensen,~W.~L.; Maxwell,~D.~S.; Tirado-Rives,~J. Development and Testing of
  the OPLS All-Atom Force Field on Conformational Energetics and Properties of
  Organic Liquids. \emph{Journal of the American Chemical Society}
  \textbf{1996}, \emph{118}, 11225--11236\relax
\mciteBstWouldAddEndPuncttrue
\mciteSetBstMidEndSepPunct{\mcitedefaultmidpunct}
{\mcitedefaultendpunct}{\mcitedefaultseppunct}\relax
\EndOfBibitem
\bibitem[Oostenbrink \latin{et~al.}(2004)Oostenbrink, Villa, Mark, and
  Van~Gunsteren]{gromosff}
Oostenbrink,~C.; Villa,~A.; Mark,~A.~E.; Van~Gunsteren,~W.~F. A biomolecular
  force field based on the free enthalpy of hydration and solvation: The GROMOS
  force-field parameter sets 53A5 and 53A6. \emph{Journal of Computational
  Chemistry} \textbf{2004}, \emph{25}, 1656--1676\relax
\mciteBstWouldAddEndPuncttrue
\mciteSetBstMidEndSepPunct{\mcitedefaultmidpunct}
{\mcitedefaultendpunct}{\mcitedefaultseppunct}\relax
\EndOfBibitem
\bibitem[Shi \latin{et~al.}(2015)Shi, Ren, Schnieders, and
  Piquemal]{reviewcompchem}
Shi,~Y.; Ren,~P.; Schnieders,~M.; Piquemal,~J.-P. \emph{Reviews in
  Computational Chemistry Volume 28}; John Wiley and Sons, Ltd, 2015; Chapter
  2, pp 51--86\relax
\mciteBstWouldAddEndPuncttrue
\mciteSetBstMidEndSepPunct{\mcitedefaultmidpunct}
{\mcitedefaultendpunct}{\mcitedefaultseppunct}\relax
\EndOfBibitem
\bibitem[Jing \latin{et~al.}(2019)Jing, Liu, Cheng, Qi, Walker, Piquemal, and
  Ren]{annurev-biophys-070317-033349}
Jing,~Z.; Liu,~C.; Cheng,~S.~Y.; Qi,~R.; Walker,~B.~D.; Piquemal,~J.-P.;
  Ren,~P. Polarizable Force Fields for Biomolecular Simulations: Recent
  Advances and Applications. \emph{Ann. Rev. Biophys.} \textbf{2019},
  \emph{48}, 371--394\relax
\mciteBstWouldAddEndPuncttrue
\mciteSetBstMidEndSepPunct{\mcitedefaultmidpunct}
{\mcitedefaultendpunct}{\mcitedefaultseppunct}\relax
\EndOfBibitem
\bibitem[Melcr and Piquemal(2019)Melcr, and Piquemal]{melcr2019accurate}
Melcr,~J.; Piquemal,~J.-P. Accurate biomolecular simulations account for
  electronic polarization. \emph{Front. Mol. Biosci.} \textbf{2019}, \emph{6},
  143\relax
\mciteBstWouldAddEndPuncttrue
\mciteSetBstMidEndSepPunct{\mcitedefaultmidpunct}
{\mcitedefaultendpunct}{\mcitedefaultseppunct}\relax
\EndOfBibitem
\bibitem[Bedrov \latin{et~al.}(2019)Bedrov, Piquemal, Borodin, MacKerell, Roux,
  and Schröder]{chemrevION}
Bedrov,~D.; Piquemal,~J.-P.; Borodin,~O.; MacKerell,~A.~D.; Roux,~B.;
  Schröder,~C. Molecular Dynamics Simulations of Ionic Liquids and
  Electrolytes Using Polarizable Force Fields. \emph{Chemical Reviews}
  \textbf{2019}, \emph{119}, 7940--7995, PMID: 31141351\relax
\mciteBstWouldAddEndPuncttrue
\mciteSetBstMidEndSepPunct{\mcitedefaultmidpunct}
{\mcitedefaultendpunct}{\mcitedefaultseppunct}\relax
\EndOfBibitem
\bibitem[Lopes \latin{et~al.}(2013)Lopes, Huang, Shim, Luo, Li, Roux, and
  MacKerell]{Drudeprot}
Lopes,~P. E.~M.; Huang,~J.; Shim,~J.; Luo,~Y.; Li,~H.; Roux,~B.;
  MacKerell,~A.~D. Polarizable Force Field for Peptides and Proteins Based on
  the Classical Drude Oscillator. \emph{Journal of Chemical Theory and
  Computation} \textbf{2013}, \emph{9}, 5430--5449, PMID: 24459460\relax
\mciteBstWouldAddEndPuncttrue
\mciteSetBstMidEndSepPunct{\mcitedefaultmidpunct}
{\mcitedefaultendpunct}{\mcitedefaultseppunct}\relax
\EndOfBibitem
\bibitem[Lemkul \latin{et~al.}(2016)Lemkul, Huang, Roux, and
  MacKerell]{drudereview}
Lemkul,~J.~A.; Huang,~J.; Roux,~B.; MacKerell,~A.~D. An Empirical Polarizable
  Force Field Based on the Classical Drude Oscillator Model: Development
  History and Recent Applications. \emph{Chemical Reviews} \textbf{2016},
  \emph{116}, 4983--5013, PMID: 26815602\relax
\mciteBstWouldAddEndPuncttrue
\mciteSetBstMidEndSepPunct{\mcitedefaultmidpunct}
{\mcitedefaultendpunct}{\mcitedefaultseppunct}\relax
\EndOfBibitem
\bibitem[Lin \latin{et~al.}(2020)Lin, Huang, Pandey, Rupakheti, Li, Roux, and
  MacKerell]{Drude2020}
Lin,~F.-Y.; Huang,~J.; Pandey,~P.; Rupakheti,~C.; Li,~J.; Roux,~B.;
  MacKerell,~A.~D. Further Optimization and Validation of the Classical Drude
  Polarizable Protein Force Field. \emph{Journal of Chemical Theory and
  Computation} \textbf{2020}, \emph{16}, 3221--3239, PMID: 32282198\relax
\mciteBstWouldAddEndPuncttrue
\mciteSetBstMidEndSepPunct{\mcitedefaultmidpunct}
{\mcitedefaultendpunct}{\mcitedefaultseppunct}\relax
\EndOfBibitem
\bibitem[Ren and Ponder(2003)Ren, and Ponder]{ren2003polarizable}
Ren,~P.~Y.; Ponder,~J.~W. "Polarizable Atomic Multipole Water Model for
  Molecular Mechanics Simulation". \emph{J. Phys. Chem.} \textbf{2003},
  \emph{107}, 5933--5947\relax
\mciteBstWouldAddEndPuncttrue
\mciteSetBstMidEndSepPunct{\mcitedefaultmidpunct}
{\mcitedefaultendpunct}{\mcitedefaultseppunct}\relax
\EndOfBibitem
\bibitem[Shi \latin{et~al.}(2013)Shi, Xia, Zhang, Best, Wu, Ponder, and
  Ren]{shi}
Shi,~Y.; Xia,~Z.; Zhang,~J.; Best,~R.; Wu,~C.; Ponder,~J.~W.; Ren,~P.
  Polarizable Atomic Multipole-Based {AMOEBA} Force Field for Proteins.
  \emph{J. Chem. Theory. Comput.} \textbf{2013}, \emph{9}, 4046--4063\relax
\mciteBstWouldAddEndPuncttrue
\mciteSetBstMidEndSepPunct{\mcitedefaultmidpunct}
{\mcitedefaultendpunct}{\mcitedefaultseppunct}\relax
\EndOfBibitem
\bibitem[Zhang \latin{et~al.}(2018)Zhang, Lu, Jing, Wu, Piquemal, Ponder, and
  Ren]{DNAAMOEBA}
Zhang,~C.; Lu,~C.; Jing,~Z.; Wu,~C.; Piquemal,~J.-P.; Ponder,~J.~W.; Ren,~P.
  AMOEBA Polarizable Atomic Multipole Force Field for Nucleic Acids. \emph{J.
  Chem. Theory. Comput.} \textbf{2018}, \emph{14}, 2084--2108\relax
\mciteBstWouldAddEndPuncttrue
\mciteSetBstMidEndSepPunct{\mcitedefaultmidpunct}
{\mcitedefaultendpunct}{\mcitedefaultseppunct}\relax
\EndOfBibitem
\bibitem[Jiang \latin{et~al.}(2011)Jiang, Hardy, Phillips, MacKerell, Schulten,
  and Roux]{DrudeNAMD}
Jiang,~W.; Hardy,~D.~J.; Phillips,~J.~C.; MacKerell,~A.~D.; Schulten,~K.;
  Roux,~B. High-Performance Scalable Molecular Dynamics Simulations of a
  Polarizable Force Field Based on Classical Drude Oscillators in NAMD.
  \emph{The Journal of Physical Chemistry Letters} \textbf{2011}, \emph{2},
  87--92, PMID: 21572567\relax
\mciteBstWouldAddEndPuncttrue
\mciteSetBstMidEndSepPunct{\mcitedefaultmidpunct}
{\mcitedefaultendpunct}{\mcitedefaultseppunct}\relax
\EndOfBibitem
\bibitem[Lemkul \latin{et~al.}(2015)Lemkul, Roux, van~der Spoel, and
  MacKerell~Jr.]{DrudeGROMACS}
Lemkul,~J.~A.; Roux,~B.; van~der Spoel,~D.; MacKerell~Jr.,~A.~D. Implementation
  of extended Lagrangian dynamics in GROMACS for polarizable simulations using
  the classical Drude oscillator model. \emph{Journal of Computational
  Chemistry} \textbf{2015}, \emph{36}, 1473--1479\relax
\mciteBstWouldAddEndPuncttrue
\mciteSetBstMidEndSepPunct{\mcitedefaultmidpunct}
{\mcitedefaultendpunct}{\mcitedefaultseppunct}\relax
\EndOfBibitem
\bibitem[Lagard\`{e}re \latin{et~al.}(2018)Lagard\`{e}re, Jolly, Lipparini,
  Aviat, Stamm, Jing, Harger, Torabifard, Cisneros, Schnieders, Gresh, Maday,
  Ren, Ponder, and Piquemal]{Tinker-HP}
Lagard\`{e}re,~L.; Jolly,~L.-H.; Lipparini,~F.; Aviat,~F.; Stamm,~B.;
  Jing,~Z.~F.; Harger,~M.; Torabifard,~H.; Cisneros,~G.~A.; Schnieders,~M.~J.;
  Gresh,~N.; Maday,~Y.; Ren,~P.~Y.; Ponder,~J.~W.; Piquemal,~J.-P. Tinker-{HP}:
  a massively parallel molecular dynamics package for multiscale simulations of
  large complex systems with advanced point dipole polarizable force fields.
  \emph{Chem. Sci.} \textbf{2018}, \emph{9}, 956--972\relax
\mciteBstWouldAddEndPuncttrue
\mciteSetBstMidEndSepPunct{\mcitedefaultmidpunct}
{\mcitedefaultendpunct}{\mcitedefaultseppunct}\relax
\EndOfBibitem
\bibitem[Rackers \latin{et~al.}(2018)Rackers, Wang, Lu, Laury, Lagard\`ere,
  Schnieders, Piquemal, Ren, and Ponder]{TINKER8}
Rackers,~J.~A.; Wang,~Z.; Lu,~C.; Laury,~M.~L.; Lagard\`ere,~L.;
  Schnieders,~M.~J.; Piquemal,~J.-P.; Ren,~P.; Ponder,~J.~W. Tinker 8: Software
  Tools for Molecular Design. \emph{J. Chem. Theory and Comput.} \textbf{2018},
  \emph{14}, 5273--5289\relax
\mciteBstWouldAddEndPuncttrue
\mciteSetBstMidEndSepPunct{\mcitedefaultmidpunct}
{\mcitedefaultendpunct}{\mcitedefaultseppunct}\relax
\EndOfBibitem
\bibitem[Jolly \latin{et~al.}(2019)Jolly, Duran, Lagard{\`e}re, Ponder, Ren,
  and Piquemal]{jolly2019raising}
Jolly,~L.-H.; Duran,~A.; Lagard{\`e}re,~L.; Ponder,~J.~W.; Ren,~P.;
  Piquemal,~J.-P. Raising the Performance of the Tinker-{HP} Molecular Modeling
  Package [Article v1.0]. \emph{LiveCoMS.} \textbf{2019}, \emph{1}, 10409\relax
\mciteBstWouldAddEndPuncttrue
\mciteSetBstMidEndSepPunct{\mcitedefaultmidpunct}
{\mcitedefaultendpunct}{\mcitedefaultseppunct}\relax
\EndOfBibitem
\bibitem[Stone \latin{et~al.}(2010)Stone, Hardy, Ufimtsev, and
  Schulten]{STONE2010116}
Stone,~J.~E.; Hardy,~D.~J.; Ufimtsev,~I.~S.; Schulten,~K. GPU-accelerated
  molecular modeling coming of age. \emph{Journal of Molecular Graphics and
  Modelling} \textbf{2010}, \emph{29}, 116 -- 125\relax
\mciteBstWouldAddEndPuncttrue
\mciteSetBstMidEndSepPunct{\mcitedefaultmidpunct}
{\mcitedefaultendpunct}{\mcitedefaultseppunct}\relax
\EndOfBibitem
\bibitem[Götz \latin{et~al.}(2012)Götz, Williamson, Xu, Poole, Le~Grand, and
  Walker]{routinems1}
Götz,~A.~W.; Williamson,~M.~J.; Xu,~D.; Poole,~D.; Le~Grand,~S.; Walker,~R.~C.
  Routine Microsecond Molecular Dynamics Simulations with AMBER on GPUs. 1.
  Generalized Born. \emph{Journal of Chemical Theory and Computation}
  \textbf{2012}, \emph{8}, 1542--1555, PMID: 22582031\relax
\mciteBstWouldAddEndPuncttrue
\mciteSetBstMidEndSepPunct{\mcitedefaultmidpunct}
{\mcitedefaultendpunct}{\mcitedefaultseppunct}\relax
\EndOfBibitem
\bibitem[Páll \latin{et~al.}(2020)Páll, Zhmurov, Bauer, Abraham, Lundborg,
  Gray, Hess, and Lindahl]{GROMACSGPUs}
Páll,~S.; Zhmurov,~A.; Bauer,~P.; Abraham,~M.; Lundborg,~M.; Gray,~A.;
  Hess,~B.; Lindahl,~E. Heterogeneous Parallelization and Acceleration of
  Molecular Dynamics Simulations in GROMACS. 2020\relax
\mciteBstWouldAddEndPuncttrue
\mciteSetBstMidEndSepPunct{\mcitedefaultmidpunct}
{\mcitedefaultendpunct}{\mcitedefaultseppunct}\relax
\EndOfBibitem
\bibitem[Salomon-Ferrer \latin{et~al.}(2013)Salomon-Ferrer, Götz, Poole,
  Le~Grand, and Walker]{routinems2}
Salomon-Ferrer,~R.; Götz,~A.~W.; Poole,~D.; Le~Grand,~S.; Walker,~R.~C.
  Routine Microsecond Molecular Dynamics Simulations with AMBER on GPUs. 2.
  Explicit Solvent Particle Mesh Ewald. \emph{Journal of Chemical Theory and
  Computation} \textbf{2013}, \emph{9}, 3878--3888, PMID: 26592383\relax
\mciteBstWouldAddEndPuncttrue
\mciteSetBstMidEndSepPunct{\mcitedefaultmidpunct}
{\mcitedefaultendpunct}{\mcitedefaultseppunct}\relax
\EndOfBibitem
\bibitem[Eastman \latin{et~al.}(2017)Eastman, Swails, Chodera, McGibbon, Zhao,
  Beauchamp, Wang, Simmonett, Harrigan, Stern, Wiewiora, Brooks, and
  Pande]{OpenMM7}
Eastman,~P.; Swails,~J.; Chodera,~J.~D.; McGibbon,~R.~T.; Zhao,~Y.;
  Beauchamp,~K.~A.; Wang,~L.-P.; Simmonett,~A.~C.; Harrigan,~M.~P.;
  Stern,~C.~D.; Wiewiora,~R.~P.; Brooks,~B.~R.; Pande,~V.~S. OpenMM 7: Rapid
  development of high performance algorithms for molecular dynamics. \emph{PLOS
  Computational Biology} \textbf{2017}, \emph{13}, 1--17\relax
\mciteBstWouldAddEndPuncttrue
\mciteSetBstMidEndSepPunct{\mcitedefaultmidpunct}
{\mcitedefaultendpunct}{\mcitedefaultseppunct}\relax
\EndOfBibitem
\bibitem[Harger \latin{et~al.}(2017)Harger, Li, Wang, Dalby, Lagardère,
  Piquemal, Ponder, and Ren]{TINKEROpenMM}
Harger,~M.; Li,~D.; Wang,~Z.; Dalby,~K.; Lagardère,~L.; Piquemal,~J.-P.;
  Ponder,~J.; Ren,~P. Tinker-OpenMM: Absolute and relative alchemical free
  energies using AMOEBA on GPUs. \emph{Journal of Computational Chemistry}
  \textbf{2017}, \emph{38}, 2047--2055\relax
\mciteBstWouldAddEndPuncttrue
\mciteSetBstMidEndSepPunct{\mcitedefaultmidpunct}
{\mcitedefaultendpunct}{\mcitedefaultseppunct}\relax
\EndOfBibitem
\bibitem[Potluri \latin{et~al.}(2016)Potluri, Luehr, and
  Sakharnykh]{potluri2016simplifying}
Potluri,~S.; Luehr,~N.; Sakharnykh,~N. Simplifying Multi-GPU Communication with
  NVSHMEM. GPU Technology Conference. 2016\relax
\mciteBstWouldAddEndPuncttrue
\mciteSetBstMidEndSepPunct{\mcitedefaultmidpunct}
{\mcitedefaultendpunct}{\mcitedefaultseppunct}\relax
\EndOfBibitem
\bibitem[Frenkel and Smit(2001)Frenkel, and Smit]{frenkel2001understanding}
Frenkel,~D.; Smit,~B. \emph{Understanding molecular simulation: from algorithms
  to applications}; Elsevier, 2001; Vol.~1\relax
\mciteBstWouldAddEndPuncttrue
\mciteSetBstMidEndSepPunct{\mcitedefaultmidpunct}
{\mcitedefaultendpunct}{\mcitedefaultseppunct}\relax
\EndOfBibitem
\bibitem[Lagard{\`e}re \latin{et~al.}(2018)Lagard{\`e}re, Jolly, Lipparini,
  Aviat, Stamm, Jing, Harger, Torabifard, Cisneros, Schnieders, Gresh, Maday,
  Ren, Ponder, and Piquemal]{lagardere2018tinker}
Lagard{\`e}re,~L.; Jolly,~L.~H.; Lipparini,~F.; Aviat,~F.; Stamm,~B.;
  Jing,~Z.~F.; Harger,~M.; Torabifard,~H.; Cisneros,~G.~A.; Schnieders,~M.~J.;
  Gresh,~N.; Maday,~Y.; Ren,~P.~Y.; Ponder,~J.~W.; Piquemal,~J.-P. Tinker-{HP}:
  a massively parallel molecular dynamics package for multiscale simulations of
  large complex systems with advanced point dipole polarizable force fields.
  \emph{Chem. Sci.} \textbf{2018}, \emph{9}, 956--972\relax
\mciteBstWouldAddEndPuncttrue
\mciteSetBstMidEndSepPunct{\mcitedefaultmidpunct}
{\mcitedefaultendpunct}{\mcitedefaultseppunct}\relax
\EndOfBibitem
\bibitem[Wienke \latin{et~al.}(2012)Wienke, Springer, Terboven, and
  an~Mey]{10.1007/978-3-642-32820-6_85}
Wienke,~S.; Springer,~P.; Terboven,~C.; an~Mey,~D. OpenACC --- First
  Experiences with Real-World Applications. Euro-Par 2012 Parallel Processing.
  Berlin, Heidelberg, 2012; pp 859--870\relax
\mciteBstWouldAddEndPuncttrue
\mciteSetBstMidEndSepPunct{\mcitedefaultmidpunct}
{\mcitedefaultendpunct}{\mcitedefaultseppunct}\relax
\EndOfBibitem
\bibitem[Chandrasekaran and Juckeland(2017)Chandrasekaran, and
  Juckeland]{10.5555/3175812}
Chandrasekaran,~S.; Juckeland,~G. \emph{OpenACC for Programmers: Concepts and
  Strategies}, 1st ed.; Addison-Wesley Professional, 2017\relax
\mciteBstWouldAddEndPuncttrue
\mciteSetBstMidEndSepPunct{\mcitedefaultmidpunct}
{\mcitedefaultendpunct}{\mcitedefaultseppunct}\relax
\EndOfBibitem
\bibitem[Sanders and Kandrot(2010)Sanders, and Kandrot]{sanders2010cuda}
Sanders,~J.; Kandrot,~E. \emph{CUDA by example: an introduction to
  general-purpose GPU programming}; Addison-Wesley Professional, 2010\relax
\mciteBstWouldAddEndPuncttrue
\mciteSetBstMidEndSepPunct{\mcitedefaultmidpunct}
{\mcitedefaultendpunct}{\mcitedefaultseppunct}\relax
\EndOfBibitem
\bibitem[Volkov(2016)]{Volkov:EECS-2016-143}
Volkov,~V. Understanding Latency Hiding on GPUs. Ph.D.\ thesis, EECS
  Department, University of California, Berkeley, 2016\relax
\mciteBstWouldAddEndPuncttrue
\mciteSetBstMidEndSepPunct{\mcitedefaultmidpunct}
{\mcitedefaultendpunct}{\mcitedefaultseppunct}\relax
\EndOfBibitem
\bibitem[Kraus(2013)]{kraus2013introduction}
Kraus,~J. An introduction to CUDA-aware MPI. \emph{Weblog entry]. PARALLEL
  FORALL} \textbf{2013}, \relax
\mciteBstWouldAddEndPunctfalse
\mciteSetBstMidEndSepPunct{\mcitedefaultmidpunct}
{}{\mcitedefaultseppunct}\relax
\EndOfBibitem
\bibitem[Essmann \latin{et~al.}(1995)Essmann, Perera, Berkowitz, Darden, Lee,
  and Pedersen]{SPME}
Essmann,~U.; Perera,~L.; Berkowitz,~M.~L.; Darden,~T.; Lee,~H.; Pedersen,~L.~G.
  A smooth particle mesh Ewald method. \emph{The Journal of Chemical Physics}
  \textbf{1995}, \emph{103}, 8577--8593\relax
\mciteBstWouldAddEndPuncttrue
\mciteSetBstMidEndSepPunct{\mcitedefaultmidpunct}
{\mcitedefaultendpunct}{\mcitedefaultseppunct}\relax
\EndOfBibitem
\bibitem[Lagardère \latin{et~al.}(2015)Lagardère, Lipparini, Polack, Stamm,
  Cancès, Schnieders, Ren, Maday, and Piquemal]{lagardere2015}
Lagardère,~L.; Lipparini,~F.; Polack,~E.; Stamm,~B.; Cancès,~E.;
  Schnieders,~M.; Ren,~P.; Maday,~Y.; Piquemal,~J.-P. Scalable Evaluation of
  Polarization Energy and Associated Forces in Polarizable Molecular Dynamics:
  II. Toward Massively Parallel Computations Using Smooth Particle Mesh Ewald.
  \emph{Journal of Chemical Theory and Computation} \textbf{2015}, \emph{11},
  2589--2599, PMID: 26575557\relax
\mciteBstWouldAddEndPuncttrue
\mciteSetBstMidEndSepPunct{\mcitedefaultmidpunct}
{\mcitedefaultendpunct}{\mcitedefaultseppunct}\relax
\EndOfBibitem
\bibitem[CUD(2020)]{CUDAFFT}
NVIDIA Corporation, CUDA Toolkit 11.1 CUFFT Library Programming Guide 2020
  ,http://developer.nvidia.com/nvidia-gpu-computing-documentation. 2020\relax
\mciteBstWouldAddEndPuncttrue
\mciteSetBstMidEndSepPunct{\mcitedefaultmidpunct}
{\mcitedefaultendpunct}{\mcitedefaultseppunct}\relax
\EndOfBibitem
\bibitem[Lipparini \latin{et~al.}(2014)Lipparini, Lagardère, Stamm, Cancès,
  Schnieders, Ren, Maday, and Piquemal]{lipparini2014}
Lipparini,~F.; Lagardère,~L.; Stamm,~B.; Cancès,~E.; Schnieders,~M.; Ren,~P.;
  Maday,~Y.; Piquemal,~J.-P. Scalable Evaluation of Polarization Energy and
  Associated Forces in Polarizable Molecular Dynamics: I. Toward Massively
  Parallel Direct Space Computations. \emph{Journal of Chemical Theory and
  Computation} \textbf{2014}, \emph{10}, 1638--1651, PMID: 26512230\relax
\mciteBstWouldAddEndPuncttrue
\mciteSetBstMidEndSepPunct{\mcitedefaultmidpunct}
{\mcitedefaultendpunct}{\mcitedefaultseppunct}\relax
\EndOfBibitem
\bibitem[{Phillips} \latin{et~al.}(2002){Phillips}, {Gengbin Zheng}, {Kumar},
  and {Kale}]{PhillipsFFT}
{Phillips},~J.~C.; {Gengbin Zheng},; {Kumar},~S.; {Kale},~L.~V. NAMD:
  Biomolecular Simulation on Thousands of Processors. SC '02: Proceedings of
  the 2002 ACM/IEEE Conference on Supercomputing. 2002; pp 36--36\relax
\mciteBstWouldAddEndPuncttrue
\mciteSetBstMidEndSepPunct{\mcitedefaultmidpunct}
{\mcitedefaultendpunct}{\mcitedefaultseppunct}\relax
\EndOfBibitem
\bibitem[Tuckerman \latin{et~al.}(1992)Tuckerman, Berne, and
  Martyna]{tuckerman1992reversible}
Tuckerman,~M.; Berne,~B.~J.; Martyna,~G.~J. Reversible multiple time scale
  molecular dynamics. \emph{J. Chem. Phys.} \textbf{1992}, \emph{97},
  1990--2001\relax
\mciteBstWouldAddEndPuncttrue
\mciteSetBstMidEndSepPunct{\mcitedefaultmidpunct}
{\mcitedefaultendpunct}{\mcitedefaultseppunct}\relax
\EndOfBibitem
\bibitem[Bussi \latin{et~al.}(2007)Bussi, Donadio, and Parrinello]{Bussi}
Bussi,~G.; Donadio,~D.; Parrinello,~M. Canonical sampling through velocity
  rescaling. \emph{J. Chem. Phys.} \textbf{2007}, \emph{126}, 014101\relax
\mciteBstWouldAddEndPuncttrue
\mciteSetBstMidEndSepPunct{\mcitedefaultmidpunct}
{\mcitedefaultendpunct}{\mcitedefaultseppunct}\relax
\EndOfBibitem
\bibitem[Zhou and Ross(2002)Zhou, and Ross]{zhou2002implementing}
Zhou,~J.; Ross,~K.~A. Implementing database operations using SIMD instructions.
  Proceedings of the 2002 ACM SIGMOD international conference on Management of
  data. 2002; pp 145--156\relax
\mciteBstWouldAddEndPuncttrue
\mciteSetBstMidEndSepPunct{\mcitedefaultmidpunct}
{\mcitedefaultendpunct}{\mcitedefaultseppunct}\relax
\EndOfBibitem
\bibitem[Nickolls and Dally(2010)Nickolls, and Dally]{nickolls2010gpu}
Nickolls,~J.; Dally,~W.~J. The GPU computing era. \emph{IEEE micro}
  \textbf{2010}, \emph{30}, 56--69\relax
\mciteBstWouldAddEndPuncttrue
\mciteSetBstMidEndSepPunct{\mcitedefaultmidpunct}
{\mcitedefaultendpunct}{\mcitedefaultseppunct}\relax
\EndOfBibitem
\bibitem[{Le Grand} \latin{et~al.}(2013){Le Grand}, Götz, and
  Walker]{LEGRAND2013374}
{Le Grand},~S.; Götz,~A.~W.; Walker,~R.~C. SPFP: Speed without compromise—A
  mixed precision model for GPU accelerated molecular dynamics simulations.
  \emph{Computer Physics Communications} \textbf{2013}, \emph{184}, 374 --
  380\relax
\mciteBstWouldAddEndPuncttrue
\mciteSetBstMidEndSepPunct{\mcitedefaultmidpunct}
{\mcitedefaultendpunct}{\mcitedefaultseppunct}\relax
\EndOfBibitem
\bibitem[Yatepep(2009)]{yates2009fixed}
Yatepep,~R. Fixed-point arithmetic: An introduction. \emph{Digital Signal Labs}
  \textbf{2009}, \emph{81}, 198\relax
\mciteBstWouldAddEndPuncttrue
\mciteSetBstMidEndSepPunct{\mcitedefaultmidpunct}
{\mcitedefaultendpunct}{\mcitedefaultseppunct}\relax
\EndOfBibitem
\bibitem[Bowers \latin{et~al.}(2006)Bowers, Dror, and Shaw]{midpoint}
Bowers,~K.~J.; Dror,~R.~O.; Shaw,~D.~E. The midpoint method for parallelization
  of particle simulations. \emph{The Journal of Chemical Physics}
  \textbf{2006}, \emph{124}, 184109\relax
\mciteBstWouldAddEndPuncttrue
\mciteSetBstMidEndSepPunct{\mcitedefaultmidpunct}
{\mcitedefaultendpunct}{\mcitedefaultseppunct}\relax
\EndOfBibitem
\bibitem[Lagardère \latin{et~al.}(2019)Lagardère, Aviat, and
  Piquemal]{pushing}
Lagardère,~L.; Aviat,~F.; Piquemal,~J.-P. Pushing the Limits of
  Multiple-Time-Step Strategies for Polarizable Point Dipole Molecular
  Dynamics. \emph{The Journal of Physical Chemistry Letters} \textbf{2019},
  \emph{10}, 2593--2599\relax
\mciteBstWouldAddEndPuncttrue
\mciteSetBstMidEndSepPunct{\mcitedefaultmidpunct}
{\mcitedefaultendpunct}{\mcitedefaultseppunct}\relax
\EndOfBibitem
\bibitem[Célerse \latin{et~al.}(2019)Célerse, Lagard{\`e}re, Derat, and
  Piquemal]{Celerse2019}
Célerse,~F.; Lagard{\`e}re,~L.; Derat,~E.; Piquemal,~J.-P. Massively parallel
  implementation of {S}teered {M}olecular {D}ynamics in
  {T}inker-{HP}:comparisons of polarizable and non-polarizable simulations of
  realistic systems. \emph{J. Chem. Theory. Comput.} \textbf{2019}, \emph{15},
  3694--3709\relax
\mciteBstWouldAddEndPuncttrue
\mciteSetBstMidEndSepPunct{\mcitedefaultmidpunct}
{\mcitedefaultendpunct}{\mcitedefaultseppunct}\relax
\EndOfBibitem
\bibitem[Miao \latin{et~al.}(2015)Miao, Feher, and McCammon]{miao2015gaussian}
Miao,~Y.; Feher,~V.~A.; McCammon,~J.~A. Gaussian accelerated molecular
  dynamics: Unconstrained enhanced sampling and free energy calculation.
  \emph{Journal of chemical theory and computation} \textbf{2015}, \emph{11},
  3584--3595\relax
\mciteBstWouldAddEndPuncttrue
\mciteSetBstMidEndSepPunct{\mcitedefaultmidpunct}
{\mcitedefaultendpunct}{\mcitedefaultseppunct}\relax
\EndOfBibitem
\bibitem[Jorgensen \latin{et~al.}(1996)Jorgensen, Maxwell, and
  Tirado-Rives]{jorgensen1996development}
Jorgensen,~W.~L.; Maxwell,~D.~S.; Tirado-Rives,~J. "Development and Testing of
  the OPLS All-Atom Force Field on Conformational Energetics and Properties of
  Organic Liquids". \emph{J. Am. Chem. Soc.} \textbf{1996}, \emph{117},
  11225--11236\relax
\mciteBstWouldAddEndPuncttrue
\mciteSetBstMidEndSepPunct{\mcitedefaultmidpunct}
{\mcitedefaultendpunct}{\mcitedefaultseppunct}\relax
\EndOfBibitem
\bibitem[Jaffrelot-Inizan \latin{et~al.}(2021)Jaffrelot-Inizan, C{\'e}lerse,
  Adjoua, El~Ahdab, Jolly, Liu, Ren, Montes, Lagarde, Lagard{\`e}re,
  \latin{et~al.} others]{JaffrelotInizan2020}
Jaffrelot-Inizan,~T.; C{\'e}lerse,~F.; Adjoua,~O.; El~Ahdab,~D.; Jolly,~L.-H.;
  Liu,~C.; Ren,~P.; Montes,~M.; Lagarde,~N.; Lagard{\`e}re,~L., \latin{et~al.}
  High-Resolution Mining of SARS-CoV-2 Main Protease Conformational Space:
  Supercomputer-Driven Unsupervised Adaptive Sampling. \emph{Chemical Science}
  \textbf{2021}, \relax
\mciteBstWouldAddEndPunctfalse
\mciteSetBstMidEndSepPunct{\mcitedefaultmidpunct}
{}{\mcitedefaultseppunct}\relax
\EndOfBibitem
\bibitem[Chapman \latin{et~al.}(2010)Chapman, Curtis, Pophale, Poole, Kuehn,
  Koelbel, and Smith]{chapman2010introducing}
Chapman,~B.; Curtis,~T.; Pophale,~S.; Poole,~S.; Kuehn,~J.; Koelbel,~C.;
  Smith,~L. Introducing OpenSHMEM: SHMEM for the PGAS community. Proceedings of
  the Fourth Conference on Partitioned Global Address Space Programming Model.
  2010; pp 1--3\relax
\mciteBstWouldAddEndPuncttrue
\mciteSetBstMidEndSepPunct{\mcitedefaultmidpunct}
{\mcitedefaultendpunct}{\mcitedefaultseppunct}\relax
\EndOfBibitem
\bibitem[Liu \latin{et~al.}(2019)Liu, Piquemal, and Ren]{CW1}
Liu,~C.; Piquemal,~J.-P.; Ren,~P. AMOEBA+ Classical Potential for Modeling
  Molecular Interactions. \emph{Journal of Chemical Theory and Computation}
  \textbf{2019}, \emph{15}, 4122--4139, PMID: 31136175\relax
\mciteBstWouldAddEndPuncttrue
\mciteSetBstMidEndSepPunct{\mcitedefaultmidpunct}
{\mcitedefaultendpunct}{\mcitedefaultseppunct}\relax
\EndOfBibitem
\bibitem[Liu \latin{et~al.}(2020)Liu, Piquemal, and Ren]{CW2}
Liu,~C.; Piquemal,~J.-P.; Ren,~P. Implementation of Geometry-Dependent Charge
  Flux into the Polarizable AMOEBA+ Potential. \emph{The Journal of Physical
  Chemistry Letters} \textbf{2020}, \emph{11}, 419--426, PMID: 31865706\relax
\mciteBstWouldAddEndPuncttrue
\mciteSetBstMidEndSepPunct{\mcitedefaultmidpunct}
{\mcitedefaultendpunct}{\mcitedefaultseppunct}\relax
\EndOfBibitem
\bibitem[Gresh \latin{et~al.}(2007)Gresh, Cisneros, Darden, and
  Piquemal]{SIBFA}
Gresh,~N.; Cisneros,~G.~A.; Darden,~T.~A.; Piquemal,~J.-P. \emph{Journal of
  Chemical Theory and Computation} \textbf{2007}, \emph{3}, 1960--1986, PMID:
  18978934\relax
\mciteBstWouldAddEndPuncttrue
\mciteSetBstMidEndSepPunct{\mcitedefaultmidpunct}
{\mcitedefaultendpunct}{\mcitedefaultseppunct}\relax
\EndOfBibitem
\bibitem[Bonomi \latin{et~al.}(2019)Bonomi, Bussi, Camilloni, Tribello,
  Ban{\'a}{\v{s}}, Barducci, Bernetti, Bolhuis, Bottaro, Branduardi,
  \latin{et~al.} others]{bonomi2019promoting}
Bonomi,~M.; Bussi,~G.; Camilloni,~C.; Tribello,~G.~A.; Ban{\'a}{\v{s}},~P.;
  Barducci,~A.; Bernetti,~M.; Bolhuis,~P.~G.; Bottaro,~S.; Branduardi,~D.,
  \latin{et~al.}  Promoting transparency and reproducibility in enhanced
  molecular simulations. \emph{Nature methods} \textbf{2019}, \emph{16},
  670--673\relax
\mciteBstWouldAddEndPuncttrue
\mciteSetBstMidEndSepPunct{\mcitedefaultmidpunct}
{\mcitedefaultendpunct}{\mcitedefaultseppunct}\relax
\EndOfBibitem
\end{mcitethebibliography}

\newpage

\makeatletter
\setlength\acs@tocentry@height{16.0cm}
\setlength\acs@tocentry@width{9.0cm}
\makeatother

\begin{tocentry}
    \begin{center}
    \includegraphics[width=\linewidth]{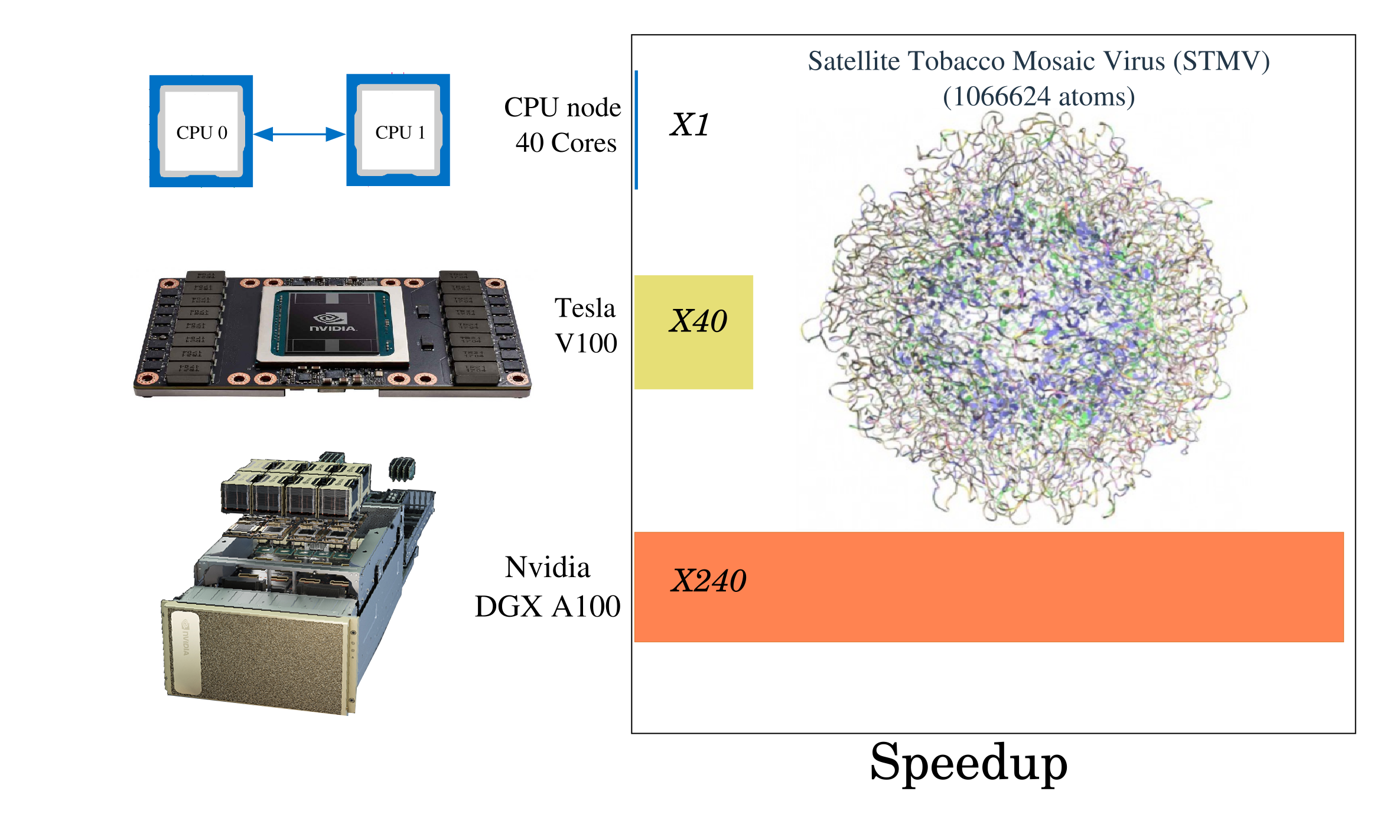}
    \end{center}
\end{tocentry}

\end{document}



%
\section{non PFF performance}
The present package also proposes non-polarizable force fields simlation capabilities. The following Table displays the performances of the initial portage of the Tinker-HP CPUs subroutines for force fields like CHARMM, AMBER or OPLS (same scaling). We are currently working on the HPC optimization of these approaches but the present implementation already allow to perform production simulations.
\begin{table}[h]
    \centering
    \begin{tabular}{c|c c c|}
                & \dhfr & \cox  & \stmv \\ \hline
    V100 (ns/j) &   130 & 28.74 & 4.71  \\
    RTX 2080 Ti &  95.5 & 14.34 & 2.37  \\ \hline
    \end{tabular}
    \caption{\TinkerHP Mixed Precision benchmarks (standard setup) with the Charmm Forcefield on a single device with RESPA \SI{2}{fs} (full flexibility, no shake}
    \label{tab:charm_perfs}
\end{table}
\section{Comparison between device architectures}
Given the device specifications, the overall computation analysis made with the Nvidia Profiler indicates a  peak performance of 9, 8 and 5\% respectively from GV100, RTX-2080 Ti and RTX 3090 after measuring the total number of Flops and duration on a 100 steps \MP simulation. In other words, each of these devices performs on average \num{6} \num{7} and \num{8} floating point operations after every access to a {32}{bits} floating point from global memory. Thus, it remains possible to add extra computational instructions which is encouraging regarding future developments. For the current generation of Geforce \GPU, it seems that reaching peak performance will represent a considerable challenge as memory technologies do not evolve proportionally as processors.